\documentclass[preprint,12pt]{elsarticle}

\newtheorem{definition}{Definition}
\newtheorem{remark}[definition]{Remark}

\newcommand{\prob}{\mathbb{P}}
\newcommand{\sgn}{\text{sgn}}
\newcommand{\expect}{\mathbb{E}}

\usepackage{amsmath,amssymb,booktabs,bm,commath,mathtools,graphicx,epstopdf}

\journal{Journal of Computational and Applied Mathematics}

\begin{document}

\begin{frontmatter}



\title{Jump Diffusion and $\alpha$-Stable Techniques for the Markov Switching Approach to Financial Time Series}


\author{Luca Di Persio\corref{cor1}} 
\ead{luca.dipersio@univr.it}

\author{Vukasin Jovic\corref{cor2}}
\ead{jovic\_vukasin@yahoo.com}

\address{University of Verona, Department of Computer Science, Strada Le Grazie 15, 37100 Verona, Italy}

\cortext[cor1]{Corresponding author}

\begin{abstract}
We perform a  detailed comparison between a \emph{Markov Switching Jump Diffusion Model} and a \emph{Markov Switching $\alpha$-Stable Distribution Model} with respect to the analysis of non-stationary data. We show that the jump-diffusion model is extremely robust, flexible and accurate in fitting of financial time series. A thorough computational study involving the two models being applied to real data, namely, the S\&P500 index, is provided. The study shows that the jump-diffusion model solves the over-smoothing issue stated in~\cite{DiPMF16}, while the $\alpha$-stable distribution approach is a good compromise between computational effort and performance in the estimate of implied volatility, which is a major problem widely underlined in the dedicated literature, see, e.g., ~\cite{DiPMF16}.
\end{abstract}


\end{frontmatter}


\section{Introduction}
Financial time series are hard to model, since they are heavily influenced by unpredictable events.
Natural disasters, uncertainty about public behaviour, statements from governments and central banks, etc., are all events that can drastically affect the market. As a consequence financial data do not behave the same at all times, hence we cannot assume any stationarity property for them. The latter implies that classic techniques used to analyse time series are widely inadequate to model such data, therefore alternative methods have to be developed. The family of \emph{Markov Switching Models} (MSM) constitutes a possible solution, since these models allow us to effectively address the non-stationarity of financial data.

The main idea behind the MSM is that, in order to take into account  the changes in the behaviour of the data, we allow the distribution of the observations to change over time. A general MSM model can be written in the following form
\begin{equation}\label{generalMSM}
\begin{cases}
y_t = f(S_t,\theta,\psi_{t-1})\\
S_t = g\left(\tilde{S}_{t-1},\psi_{t-1}\right) \\
S_t \in \Lambda
\end{cases}
\end{equation}
where $S_{t}$ indicates the state of the model at time $t$, $\theta$ is the vector of the parameters characterizing the model, $\psi_t:=\left\{y_k: k=1,\dots,t\right\}$ is the set of all observations up to time $t$, $\tilde{S}_{t} := \{S_{1},...,S_{t}\}$ is the set of all observed states up to time $t$, $\Lambda=\{1,...,M\}$ is the set of all possible states, and $g$ is the function that governs the transitions between the states. The function $f$ defines how the observation at time $t$ depends on $S_t,\theta,\ \text{and}\ \psi_{t-1}$ and finally, $t \in \{0,1,...,T\}$, where $T \in \mathbb{N}$, $T < +\infty$, is the so called \emph{terminal time}.
System \eqref{generalMSM} clearly shows the intrinsic richness of the MSM approach. Particular realizations of \eqref{generalMSM} allow  the treatment of specific problems. Before getting into the details of our study, it is worth mentioning that in most of the dedicated  literature, we can distinguish between two classes of models. The first class consists of models that have complicated distributions for the data or a large number of states, but very simple transition laws, e.g.,  a first order Markov chain, see, e.g., \cite{DiPMF16,Ham89,Ch98}. The second class is made up of models with simple assumptions and very few states, usually two, but with more complicated transition laws, see, e.g., \cite{Kim99, DiLeeW94, Pe05}. 

The present paper is structured as follows: in Sections 1 through 4 we cover the mathematical and computational theory needed to establish the techniques that we then develop in subsequent sections; in Section 5 we introduce the jump-diffusion model, while in Section 6 we present a model that uses $\alpha-$stable distributions;  in Section 7 we explain how the models can be concretely implemented and, in Section 8, we present the related results obtained applying them to a relevant case study which concerns the S\&P500 index; conclusions and further developments are outlined in Section 9.

\section{Bayesian Inference}
Bayesian Inference is a branch of statistical inference that assumes the parameters of a probability distribution to be randomly distributed according to a {\it prior} distribution. In particular the idea is to exploit the observed data, along with the Bayes rule, to generate the {\it posterior} distribution of the aforementioned parameters. Therefore, the posterior distribution can be interpreted as the distribution of the parameters once we have taken into account both our subjective belief about them, namely the prior, and the data. Such an approach can be rigorously represented as follows
\begin{align}
& \nonumber \theta \sim \pi(\theta) \,;\\
& y|\theta \sim f(y|\theta) \,;\\
& \nonumber f(\theta|y) = \frac{\pi(\theta)f(y|\theta)}{f(y)} \, ,
\end{align}
where $\pi(\theta)$ is the prior distribution, $f(y|\theta)$ is the distribution of the data depending on the parameter $\theta$, and $f(\theta|y)$ is the posterior of $\theta$. Finally, $f(y)$ is the marginal distribution of $y$, namely
\begin{equation*}
f(y) = \int f(y,\theta) d\theta = \int \pi(\theta)f(y|\theta) d\theta\:.
\end{equation*}
Clearly the choice of the prior can have a large impact on the posterior. A particularly convenient form of prior is what is known as a \emph{conjugate prior}. We say that a prior distribution is conjugate if the posterior distribution derived from it belongs to the same family, as it happens, e.g., for the Beta-Bernoulli pair, namely
\begin{equation}
\begin{aligned}
\pi(\theta) &= \theta^{\alpha-1}  (1-\theta)^{\beta-1} \,;\\ 
f(y|\theta) &= \binom{n}{y} \theta^{y} (1-\theta)^{n-y} \,;\\
f(\theta|y) & \propto \theta^{(\alpha + y)-1} (1-\theta)^{(\beta+n-y)-1} \,;\\
& \propto \text{Beta}(\alpha + y,\beta+n-y)\,.
\end{aligned}
\end{equation}

It follows that if we start with a $\text{Beta}(\alpha, \beta)$ prior and assume that the data are binomially distributed, we end up with a $\text{Beta}(\alpha + y,\beta+n-y)$ posterior. Hence, we do not have to update the distribution for each new observation, just its parameters. We would like to note that the latter is a particularly relevant aspect from the algorithmic point of view since it translates into less computationally expensive code. For the sake of completeness, in the following subsections we list other particularly convenient choices for distribution pairs and, in order to give clear examples, we first start by explaining how the posterior of a set of independent identically distributed (i.i.d) random variables is obtained. Let $\{y_{1},...,y_{n}\},\ n \in \mathbb{N}$ be a set of i.i.d. random variables with  
density function $f(y|\theta)$. Moreover, let $\theta \sim \pi(\theta)$. Then
\begin{equation}
f(\theta|y_{1},...,y_{n}) = \frac{\pi(\theta) L_{n}(\theta)}{\int\limits_{\Theta} \pi(\theta) L_{n}(\theta) d\theta} \propto \pi(\theta) L_{n}(\theta)\;,
\end{equation}
where $$L_{n}(\theta) = \prod\limits_{i=1}^{n} f(y_{i}|\theta)\;,$$ is the likelihood function of the data, and $\Theta$ is the set of all possible values of $\theta$. For the rest of this paper we will denote the vector of observations by $\boldsymbol y$. 

\subsection{Normal-Normal}
\label{nn}
Assume that we have  $n \in \mathbb{N}$ independent observations
\begin{align*}
y_{i}| \mu \sim \mathcal{N}(\mu,\sigma^{2}),\ i \in \{1,...,n\} \;,
\end{align*}
where $\mu \in \mathbb{R}$, and the value of $\sigma>0$ is known. In order to perform Bayesian inference on the given data, we also need to place a distribution on $\mu$. Therefore, we set
\begin{align*}
\mu &\sim \mathcal{N} \left(\mu_{0},\frac{1}{k}\right)\:.
\end{align*}
The corresponding likelihood function is
\begin{align*}
L_{n}(\mu) = \prod\limits_{i=1}^{n} \frac{1}{\sqrt{2 \pi \sigma^{2}}} e^{- \frac{1}{2} \left( \frac{y_{i}-\mu}{\sigma} \right)^{2}} &\propto e^{- \frac{1}{2\sigma^{2}} \sum\limits_{i=1}^{n} \left( y_{i}-\mu \right)^{2}} \\
\propto e^{- \frac{1}{2\sigma^{2}} \left( \sum\limits_{i=1}^{n} (y_{i}-\bar{y})^{2} + n(\bar{y}-\mu)^{2} \right)} &\propto e^{- \frac{n}{2\sigma^{2}}(\bar{y}-\mu)^{2}}
\:,
\end{align*}
while for the posterior we have
\begin{align*}
& f(\mu|\boldsymbol y) \propto e^{- \frac{n}{2\sigma^{2}}(\bar{y}-\mu)^{2}} e^{- \frac{k}{2} \left( \mu-\mu_{0} \right)^{2}} \\ 
& = e^{- \frac{1}{2} \left( \tfrac{n}{\sigma^{2}}(\bar{y}-\mu)^{2} + k(\mu-\mu_{0})^{2} \right) } \\
&= e^{-\frac{1}{2}\left( \left(\frac{n}{\sigma^{2}}+k\right) \left( \mu - \frac{\frac{n}{\sigma^{2}}\bar{y} + k \mu_{0}}{\frac{n}{\sigma^{2}}+k} \right)^2 + \frac{\frac{nk}{\sigma^{2}}}{\frac{n}{\sigma^{2}}+k} (\bar{y}-\mu_{0})^{2} \right)  } \\
& \propto \mathcal{N}\left(\frac{\frac{n}{\sigma^{2}}\bar{y} + k \mu_{0}}{\frac{n}{\sigma^{2}}+k}, \frac{\sigma^{2}}{n+k \sigma^{2}}\right)\;,
\end{align*}
hence 
\begin{equation}
\boldsymbol y \sim \mathcal{N} \left(\frac{n\bar{y}+\mu_{0}k\sigma^{2}}{n+k\sigma^{2}},\frac{\sigma^{2}}{n+k \sigma^{2}} \right)\:.
\end{equation} 

\subsection{Inverse Gamma-Normal}
\label{ign}
Assume again that we have  $n \in \mathbb{N}$ independent observations
\begin{align*}
y_{i}| \mu \sim \mathcal{N}(\mu,\sigma^{2}),\ i \in \{1,...,n\},\ n \in \mathbb{N}\;,
\end{align*}
but, this time, $\mu \in \mathbb{R}$ is known, while $\sigma>0$ is unknown. Taking $\sigma^{2}$ to be inverse-gamma distributed with parameters $\alpha_{0}$ and $\beta_{0}$, and denoting the distribution by $\text{inv}\Gamma(\alpha_{0},\beta_{0})$, we can write the density function of $\sigma$ as follows
\begin{equation}
f(\sigma) = \frac{\beta_{0}^{\alpha_{0}}}{\Gamma(\alpha_{0})} \left( \sigma^{2} \right)^{-\alpha_{0}-1} e^{-\frac{\beta_{0}}{\sigma^{2}}}\;,
\end{equation}
where $\Gamma(t)$ is the extension of the factorial to the set of positive real numbers, known as the {\it Gamma function}, and defined by
\begin{equation*}
\Gamma(t) := \int\limits_{0}^{\infty} x^{t-1} e^{-x} dx\:.
\end{equation*}
Therefore, the associated likelihood function is
\begin{equation*}
L_{n}(\sigma) \propto \left( \sigma^{2} \right)^{-\frac{n}{2}} e^{- \frac{1}{2\sigma^{2}} \sum\limits_{i=1}^{n} \left( y_{i}-\mu \right)^{2}}\;,
\end{equation*}
hence, the posterior is
\begin{equation}
\begin{aligned}
f(\sigma| \boldsymbol y)  \propto \left( \sigma^{2} \right)^{-\frac{n}{2}-\alpha_{0}-1} e^{- \frac{1}{2\sigma^{2}} \sum\limits_{i=1}^{n} \left( y_{i}-\mu \right)^{2}} e^{-\frac{\beta_{0}}{\sigma^{2}}} \\
\propto \text{inv}\Gamma \left(\frac{n}{2}+\alpha_{0}, \frac{1}{2} \sum\limits_{i=1}^{n} \left( y_{i}-\mu \right)^{2} + \beta_{0} \right)
\:.
\end{aligned}
\end{equation}
\begin{remark}
	Unfortunately, not all distribution pairs are as convenient as the previously mentioned ones, especially from the point of view of the parameter simulation needed by concrete computational studies. When the posterior is a well known distribution, as in the {\it normal-normal} and {\it inverse gamma-normal} cases, we can simulate the parameters using, e.g., existing R libraries. Otherwise, {\it ad hoc} sampling algorithms have to be developed. The next section addresses these problems.
\end{remark}

\section{Markov Chain Monte Carlo}
In this section, we describe two methods that will be used to sample the parameters, namely, the \emph{Gibbs Sampling Method} and the \emph{Metropolis-Hastings Algorithm}. The latter will be used in situations where the posterior distribution is non-standard, while the former will be used when the distribution can be simulated using an existing software.

\subsection{Gibbs Sampling}
Assume that we have a model with a finite number $k$ of parameters, $\boldsymbol \theta = (\theta_{1},...,\theta_{k})$, and that we want to find the full posterior distribution $f(\theta_{1},...,\theta_{k}|\boldsymbol y)$. This goal can be quite difficult to reach, since the multivariate simulation of distributions is much more tangled and computationally heavy than its univariate counterpart. The Gibbs sampling approach allows the sampling of $f(\theta_{1},...,\theta_{k}|\boldsymbol y)$, knowing only the conditional distributions $f(\theta_{i}|\theta_{1},...,\theta_{i-1},\theta_{i+1},...,\theta_{k}, \boldsymbol y), i \in \{1,...,k\}$.

Let $N$ be the number of simulations we want to perform. We assign arbitrary starting values $(\theta_{1}^{0},...,\theta_{k}^{0})$ to each of the parameters. Then, for every $j \in \{1,...,N\}$, we perform  the following steps
\begin{align*}
\textbf{Step 1:}\ & \text{Draw}\ \theta_{1}^{j}\ \text{from}\ f(\theta_{1}^{j}|\theta_{2}^{j-1},...,\theta_{k}^{j-1}, \boldsymbol y)\,; \\
\textbf{Step 2:}\ & \text{Draw}\ \theta_{2}^{j}\ \text{from}\ f(\theta_{2}^{j}|\theta_{1}^{j}, \theta_{3}^{j-1},...,\theta_{k}^{j-1}, \boldsymbol y)\,; \\
\vdots \\
\textbf{Step k:}\ & \text{Draw}\ \theta_{k}^{j}\ \text{from}\ f(\theta_{k}^{j}|\theta_{1}^{j},...,\theta_{k-1}^{j}, \boldsymbol y)\,;
\end{align*}
hence we can simulate each of the model parameters. The first $J$ simulations are discarded, being part of what is called the \emph{burn in period}, in order to get rid of the dependence on the arbitrary choice of the starting point $(\theta_{1}^{0},...,\theta_{k}^{0})$, while the remaining $N-J$ values are assumed to be a suitable approximation of the real distribution. It is worth mentioning that the number of iterations, as well as the length of the {\it burn in period}, should be chosen carefully, since for larger values of $N$ the simulations become too time consuming, while small values might not provide enough iterations for the sampler to converge.

\subsection{Metropolis-Hastings Algorithm}
\label{sbsec:metropolis_hastings}
The Gibbs sampler is rather easy to implement, but its major drawback is that it requires each $f(\theta_{i}| \boldsymbol \theta_{-i}, \boldsymbol y)$ to be readily samplable, where $\boldsymbol \theta_{-i}$ is the vector $\boldsymbol \theta \backslash \{\theta_{i}\}$. The Metropolis-Hastings algorithm allows for a solution to such an inconvenience. In particular, it only requires a function $f^{\star}(\theta_{i}|\boldsymbol \theta_{-i}, \boldsymbol y)$  proportional to the density function $f(\theta_{i}|\boldsymbol \theta_{-i}, \boldsymbol y)$, and a \emph{proposal distribution} $q(\cdot| \boldsymbol \theta)$ which denotes a proper probability density function defined on the space $\Theta$ of all possible values of $\boldsymbol \theta$.
In what follows we provide the description of the general Metropolis-Hastings algorithm, which uses the full parameter vector $\boldsymbol \theta$, as it is reported in \cite{Car}. We underline that the algorithm remains unchanged when  $\boldsymbol \theta$ is a scalar. 

Let $N$ be the number of simulations we want to perform. We assign an arbitrary starting value $\boldsymbol \theta_{0}$ to the parameter vector. Then, for every $j \in \{1,...,N\}$, we perform  the following steps
\begin{align*}
& \textbf{Step 1:}\ \text{Draw}\ \boldsymbol \theta_{new}\ \text{from}\ q(\cdot| \boldsymbol \theta^{j-1})\,; \\
& \textbf{Step 2:}\ \text{Compute the ratio}\ r =\frac{f^{\star}(\boldsymbol \theta_{new}) q(\boldsymbol \theta^{j-1}|\boldsymbol \theta^{new})}{f^{\star}(\boldsymbol \theta^{j-1})q(\boldsymbol \theta_{new}|\boldsymbol \theta^{j-1})} \,;\\
& \textbf{Step 3:}\ \text{Define}\ p = \min \left\lbrace 1,r \right\rbrace \,;\\
& \textbf{Step 4:}\ \text{Set}\ \boldsymbol \theta^{j} := 
\begin{cases}
\boldsymbol \theta^{new}, &\text{with probability}\ p \\
\boldsymbol \theta^{j-1}, &\text{with probability}\ 1-p
\end{cases}\,.
\end{align*}
Having defined a method to sample the parameters, we now have the task of simulating the states of the models we will be using. This problem is the subject of the next section.

\section{State Simulation}
\label{chap:states}
In this section our goal is to simulate the state vector $\tilde{S}_{T}$. In order to accomplish this, we first need to obtain the values $\prob(S_{1}|\tilde{y}_{1}),...,\prob(S_{T}|\tilde{y}_{T})$. We start by setting arbitrary values for the parameters, and then we use the following expression
\begin{equation}
\begin{aligned}
g(\tilde{S}_{T}|\tilde{y}_{T}) &= 
g(S_{T}|\tilde{y}_{T}) \prod_{t=1}^{T-1} g(S_{t}|S_{t+1}, \tilde{y}_{t}) \\
& = g(S_{T}|\tilde{y}_{T}) \prod_{t=1}^{T-1} g(S_{t+1}|S_{t}) g(S_{t}| \tilde{y}_{t})\:.
\end{aligned}
\end{equation}

Notice  that, $\forall t \in \{1,...,T-1\}$, we can sample from $\tilde{S}_{T}$ if we have $g(S_{t+1}|S_{t})$, which is nothing more than the transition probability from one state to another,  and $g(S_{t}| \tilde{y}_{t})$. The latter can be obtained, $\forall t \in \{1,...,T\}$, exploiting the Hamilton filter, see below.
\subsection*{Hamilton filter}
The basic Hamilton filter, see \cite{Ham89}, can be described as input-output-byproduct.

\begin{flushleft}
	\textbf{input:} $g(S_{t-1}=s_{t-1}|\tilde{y}_{t-1})\;;$
	
	\textbf{output:} $g(S_{t}=s_{t}|\tilde{y}_{t})\;;$
	
	\textbf{byproduct:} $f(y_{t}|\tilde{y}_{t-1})\:.$
\end{flushleft}
Running the Hamilton filter for $t \in \{1,...,T\}$, we get the desired values $g(S_{1}| \tilde{y}_{1}),..., g(S_{T}|\tilde{y}_{T})$, which can be used  to generate $\tilde{S}_{T}$, as described in what follows
\begin{align*}
\prob (S_{T}=i|\tilde{y}_{T}) &= \frac{g(S_{T}=i|\tilde{y}_{T})}{\sum\limits_{j=1}^{4} g(S_{T}=j|\tilde{y}_{T})}\;,
\end{align*}
such a  probability is used to draw a sample of $S_{T}$, i.e.
\begin{align*}
& \prob (S_{T-1}=i|\tilde{y}_{T-1}) = \frac{g(S_{T-1}=i|\tilde{y}_{T-1})}{\sum\limits_{j=1}^{4} g(S_{T-1}=j|\tilde{y}_{T-1})} \\ &= \frac{g(S_{T}|S_{T-1}=i) g(S_{T-1}=i| \tilde{y}_{T-1})}{\sum\limits_{j=1}^{4} g(S_{T}|S_{T-1}=j) g(S_{T-1}=j| \tilde{y}_{T-1})}\;,
\end{align*}
then, the above probability together with the previously simulated $S_{T}$, are both used to simulate $S_{T-1}$, and, proceeding iteratively, we have
\begin{align*}
&\ \ \vdots \\
\prob (S_{1}=i|\tilde{y}_{1}) & = \frac{g(S_{1}=i|\tilde{y}_{1})}{\sum\limits_{j=1}^{4} g(S_{1}=j|\tilde{y}_{1})} \\ & = \frac{g(S_{2}|S_{1}=i) g(S_{1}=i| \tilde{y}_{1})}{\sum\limits_{j=1}^{4} g(S_{2}|S_{1}=j) g(S_{1}=j| \tilde{y}_{1})}\;,
\end{align*}
therefore, 
we can simulate $S_{1}$ obtaining  the last component of $\tilde{S}_{T}$. The latter implies  that, for every $t \in \{1,...,T\}$, we know what the distribution of $y_{t}$ is, because we know what the state we are in is. In the next two sections we present the models that will be used later.

\section{Jump Diffusion Model}\label{JumpDiffusionModel}
In the paper {\it The Variation of Certain Speculative Prices}, see \cite{Mb63}, Benoit Mandelbrot draws attention to the fact that the normal distribution is inadequate when it comes to describing economic and financial data. 
\begin{figure}[!h]
	\includegraphics[width = \columnwidth, height = .2\paperheight]{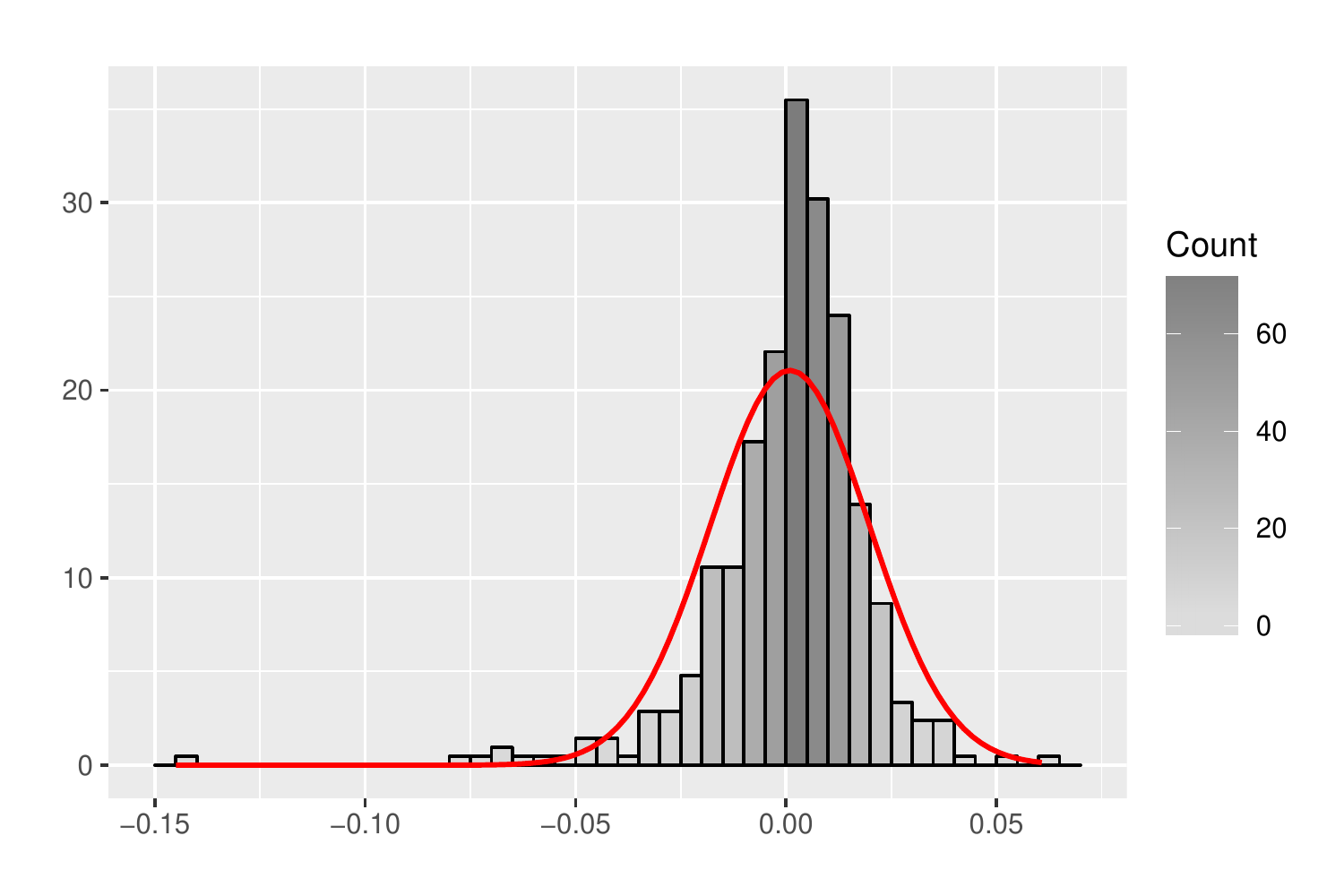}
	\caption{ \small{The histogram of the weekly log returns of the S\&P500 and the density of the normal distribution obtained by the maximum likelihood method. We see that the normal density has to {\it sacrifice} values around the mean  to cover the values at the tails.} }
\end{figure}
He argues that although the histograms of price changes seem to behave according to a Gaussian distribution,  a more careful analysis reveals that the large number of outliers makes the normal distribution fitted to the data much flatter than the actual data are, and with not enough density at the tails to include all the extreme values. If one  tries to manipulate the variance of the Gaussian distribution  to accommodate the values around the mean, then the result is a distribution that is even worse than the previous one where the extreme values are concerned. 
\begin{figure}[!h]
	\includegraphics[width = \columnwidth, height = .2\paperheight]{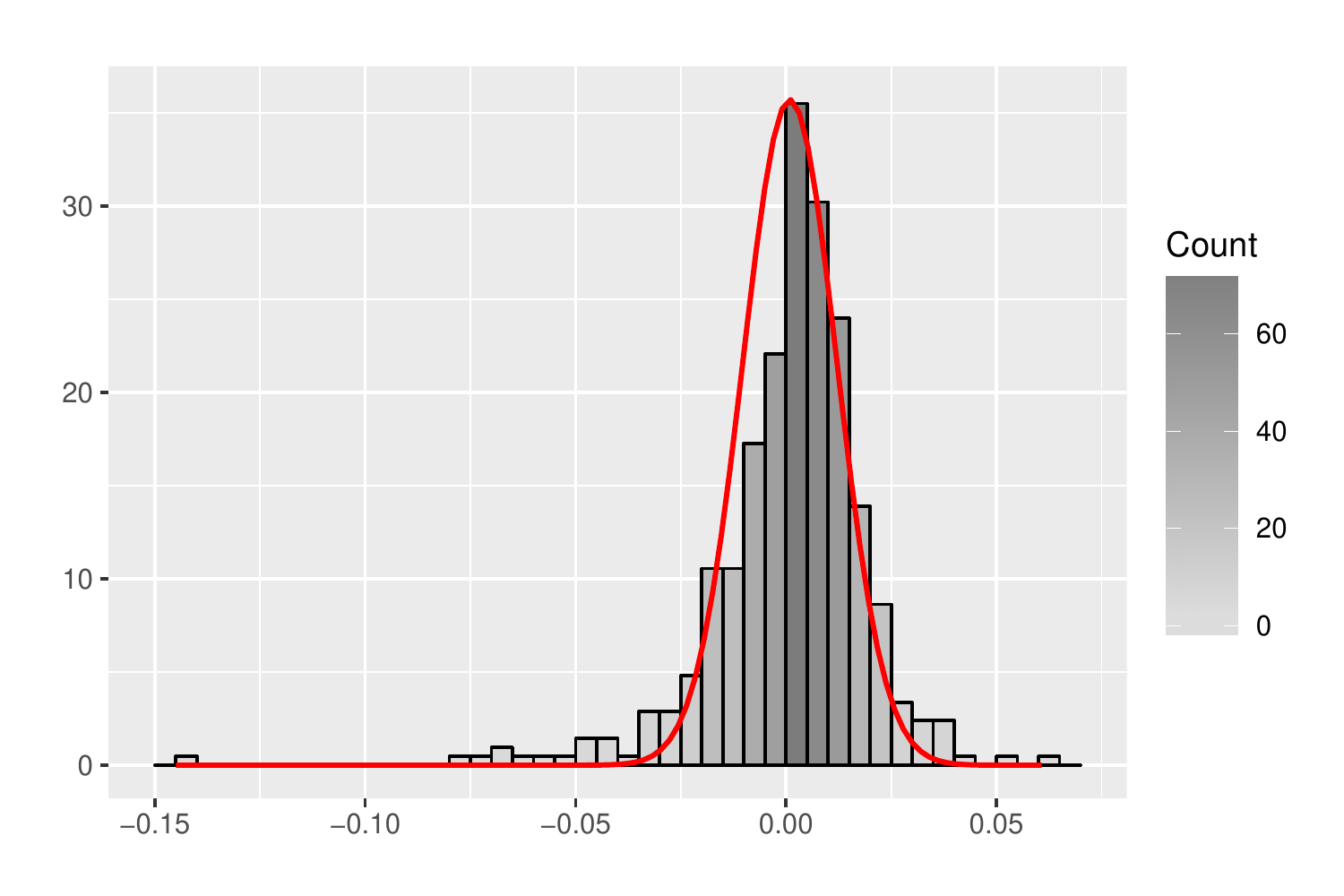}
	\caption{ \small{The histogram of the weekly log returns of the S\&P500 and the density of the normal distribution with modified variance. In this case the larger values, in absolute terms,  are severely underrepresented by the density.} }
\end{figure}
In what follows we will show how to solve the aforementioned issue by using a Gaussian distribution, to model the values around the mean, plus  jumps of stochastic intensity, to include outlying values. Specifically, our model is the following
\begin{equation}\label{model}
\begin{cases}
y_t = \epsilon_{t} + \delta \sum\limits_{i=1}^{N_{t}} z_{i} \\
\epsilon_{t} \sim \mathcal{N}(\mu_{S_{t}}, \sigma_{S_{t}}^{2}) \\
z_{i} \stackrel{i.i.d.}{\sim} \mathcal{E}(b) \\
N_{t} \sim \mathcal{P}(\theta_{S_{t}}) \\
\delta \in \{-1,1\},\ \prob (\delta=-1) = \prob (\delta=1) = \frac{1}{2} \\
\mu_{S_{t}} = \mu_{j}\ \text{if}\ S_{t}=j,\ \forall j \in \{1,...,M\} \\
\sigma_{S_{t}} = \sigma_{j}\ \text{if}\ S_{t}=j,\ \forall j \in \{1,...,M\} \\
b >0 \\
\theta_{S_{t}} = \theta_{j}\ \text{if}\ S_{t}=j,\ \forall j \in \{1,...,M\} \\
S_{t} \in \{1,...,M\} \\
p_{ij} = \prob (S_{t}=j|S_{t-1}=i) \\
\pi_{0} = [\prob (S_{0}=1),...,\prob (S_{0}=M)]
\end{cases}
\end{equation}
We divide the analysis of the model defined in \eqref{model} into two components, the {\it Gaussian component} and the {\it jump component}.

\subsection{Gaussian Element}\label{GaussianElement}
We will use the Gaussian distribution to model most of the data by means of the random variable $\epsilon_{t} \sim \mathcal{N}(\mu_{S_{t}}, \sigma_{S_{t}}^{2})$, where both the mean and the variance of $\epsilon_{t}$ are state dependent. In particular, we define the state dependence of the mean as follows
\begin{equation}
\mu_{S_{t}} = \mu_{j}\ \text{, if}\ S_{t}=j, \, \forall j \in \{1,...,M\}\;,
\end{equation}
hence each state has its own, constant mean, without further restrictions. Concerning the variance, we assume that it increases depending on the state, namely
\begin{equation}
\sigma_{S_{t}}^{2} =
\begin{cases}
\sigma^{2}_{1} , &S_{t} = 1 \\
\sigma^{2}_{1} \prod\limits_{i=2}^{S_{t}} (1+h_{i}), &S_{t} \in \{2,...,M\}
\end{cases}
\;,
\end{equation}
where, $\forall i \in \{2,3,...,M\}$, $h_{i}>0$, which gives us
\begin{equation}
\label{sigmainequality}
\sigma_{1}^{2} < \sigma_{2}^{2} < ... < \sigma_{M}^{2}\;,
\end{equation}
hence, by \eqref{sigmainequality}, as we go up in states we also go up in volatility.

\subsection{Jump Element}\label{JumpElement}
Jump diffusion models, first introduced into finance by Robert C. Merton in \cite{Mert75}, are currently widely accepted as an effective way to model the behaviour of financial data, see, e.g., \cite{Bates96, It14}. 
In order to incorporate the jump feature in our model, we have to deal with two major difficulties. First, we have to find a distribution under which the sum of independent random variables behaves well, at least from the point of view of real statistical applications. This task is not as straightforward as it may seem, since even the sum of i.i.d. uniform random variables has a distribution that rapidly grows in complexity with the number of addends. To overcome this particular problem, we have chosen to exploit the exponential distribution to model  the i.i.d. jump amplitudes, as the sum of i.i.d. exponential random variables follows a {\it Gamma distribution}, namely
\begin{equation}\label{Gamma}
z_{i} \stackrel{i.i.d.}{\sim} \mathcal{E}(b),\ 1 \leq i \leq N \Rightarrow \sum\limits_{i=1}^{N}z_{i} \sim \Gamma(N,b)\;,
\end{equation}
where $\Gamma(\alpha, \beta)$ is the Gamma distribution in the $(\alpha, \beta)$ parameterization, while $b>0$ and $N$ are given natural numbers.

The aforementioned choice leads us to the second problem, which concerns the sign of the jumps. Obviously, financial shocks  can have both positive  and negative values, while the Gamma distribution allows only for the positive ones. The issue can be solved by multiplying the  sum in \eqref{Gamma} by a random variable $\delta$ taking values in $\{-1,1\}$, with equal probability. We refer to the resulting distribution as the \emph{symmetric Gamma distribution} and we denote it by $\text{sym}\Gamma(\alpha, \beta)$. Assuming now that $X \sim \text{sym}\Gamma(\alpha, \beta)$, the probability density function of $X$ is given by
\begin{equation}
f_{X}(x) = \frac{\beta^{\alpha}}{2 \Gamma(\alpha)} |x|^{\alpha-1} e^{-\beta |x|},\ x \in \mathbb{R}\;,
\end{equation}
hence $X$ has mean equal to zero, and variance
\begin{equation}\label{GammaVar}
\begin{aligned}
\mathbb{V}(X) & = \mathbb{E}(X^{2}) - \mathbb{E}(X)^{2} = \mathbb{E}(X^{2}) \\ 
& = \int\limits_{-\infty}^{\infty} x^{2} \frac{\beta^{\alpha}}{2 \Gamma(\alpha)} |x|^{\alpha-1} e^{-\beta |x|} dx \\
& = \int\limits_{0}^{\infty} \frac{\beta^{\alpha}}{\Gamma(\alpha)} \underbrace{x^{(\alpha+2)-1} e^{-\beta x} }_{\propto \Gamma(\alpha+2, \beta)} dx \\
& = \dfrac{\beta^{\alpha}}{\Gamma(\alpha)} \frac{\Gamma(\alpha+2)}{\beta^{\alpha+2}} = \frac{\alpha(\alpha+1)}{\beta^{2}}
\end{aligned} 
\end{equation}

\begin{figure}[!h]
	\includegraphics[width = \columnwidth, height = .25\paperheight]{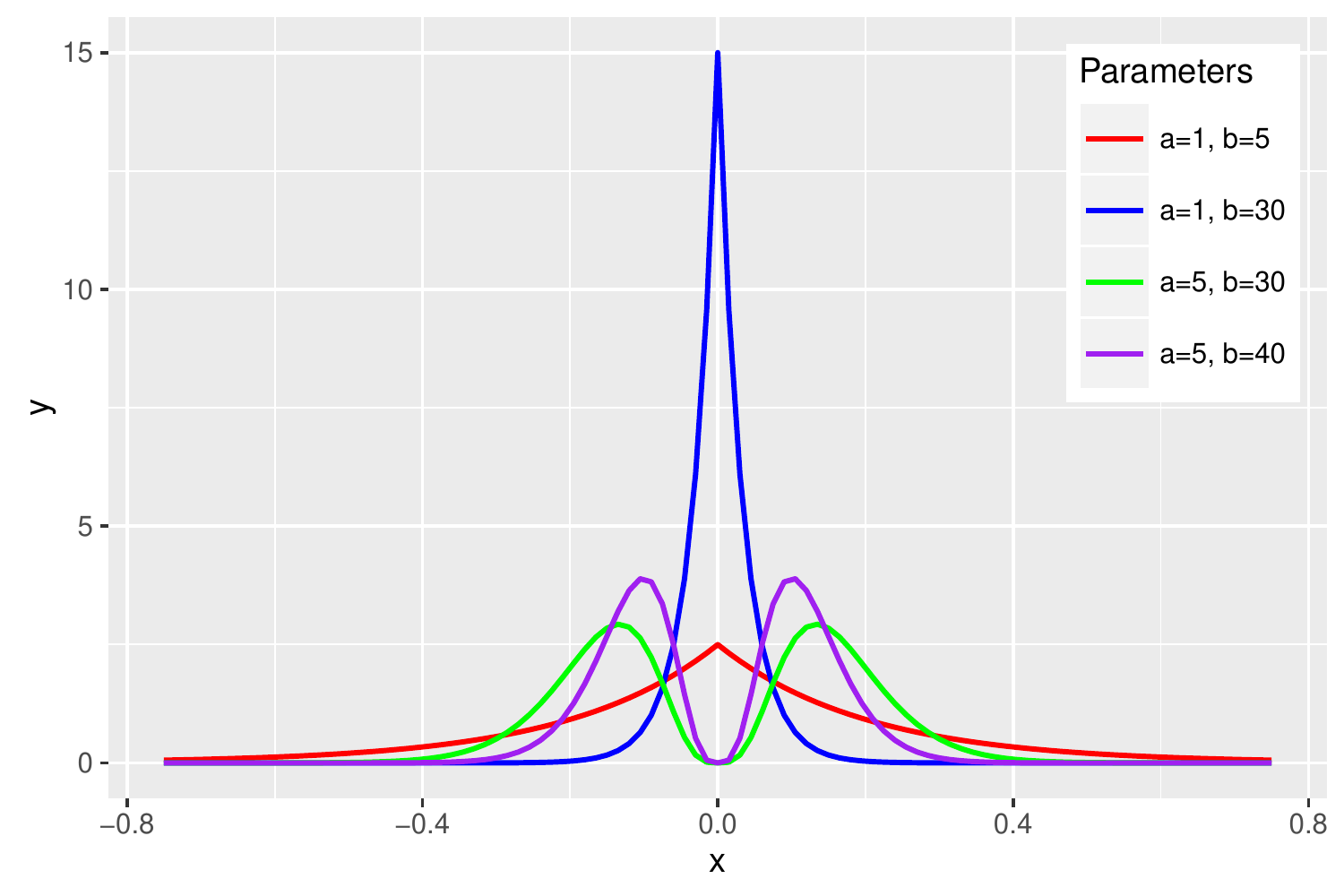}
	\caption{ \small{Comparison of four of the variants of the symmetric Gamma probability density function.} }
\end{figure}

Looking at eq. \eqref{GammaVar} one can see that $\beta$ can be used to control how much $\alpha$ influences the variance of the distribution. For example, if we take two random variables $X_{i} \sim \text{sym}\Gamma(i,1),\ i \in \{1,2\}$, we have $\mathbb{V}(X_{1})=2$ and $\mathbb{V}(X_{1})=6$ which is a drastic increase in variance. Taking $X_{i} \sim \text{sym}\Gamma(i,30),\ i \in \{1,2\}$ on the other hand gives us the variances $\mathbb{V}(X_{1}) \approx 0.0022$ and $\mathbb{V}(X_{2}) \approx 0.0066$, which is a much smaller increase, for the same change in $\alpha$. The previous, rather straightforward observation, will be useful later since in our model $\alpha$ will represent the number of jumps at a certain point in time. Taking a large $\beta$ means that every extra jump only slightly increases the variance of the model, hence  allowing for a finer analysis of the data.
The next step consists in determining the length $N$ of the sum in eq. \eqref{Gamma}. In particular we assume that such a sum has a state dependent length represented by a state-dependent Poisson random variable $N_{t} \sim \mathcal{P}(\theta_{S_{t}})$, hence we have to determine the values of $\theta_{S_{t}}$. In keeping with the interpretation of the states, see eq. \eqref{sigmainequality}, we want the number of jumps to increase as the state the data are in increases. This can be done by ordering the parameters $\theta_{S_{t}}$. Moreover, in order to also allow the parameters to be sufficiently flexible for our purposes, we assume them to be distributed as follows
\begin{equation}
\label{poissonprior}
\theta_{S_{t}} \sim 
\begin{cases}
\mathcal{U}(0,u_{1}), &{S_{t}}=1 \\
\mathcal{U}(u_{1},u_{2}), &{S_{t}}=2 \\
& \vdots \\
\mathcal{U}(u_{M-2},u_{M-1}), &{S_{t}}=M-1 \\
\mathcal{U}(u_{M-1},u_{M}), &{S_{t}}=M
\end{cases}
\end{equation}
which clearly guarantees that $\theta_{1} \leq \theta_{2} \leq ... \leq \theta_{M}$.

\subsection{Full Model}
Summing up the definitions stated in subsections \ref{GaussianElement} and \ref{JumpElement}, we can write the full model as follows
\begin{equation}
\begin{cases}
y_{t} &= \epsilon_{t} + \zeta_{t} \\
\epsilon_{t} &\sim \mathcal{N}(\mu_{S_{t}}, \sigma_{S_{t}}^{2}) \\
\zeta_{t} &\sim \text{sym}\Gamma(N_{t},b)
\end{cases}
\:.
\end{equation}
There is no analytic expression for the distribution of $y_{t}$, but we can obtain an integral form of it using the following well known fact.
Let $X$ and $Y$ be two independent random variables with density functions $f_{X}(x)$ and $f_{Y}(x)$, defined for $x \in \mathbb{R}$. Then the sum $Z = X + Y$ is a random variable with density function $f_{Z}(z)$ given by
\begin{align}\label{convolution}
f_{Z}(z) = \int\limits_{-\infty}^{\infty} f(z - y)g(y) dy = \int\limits_{-\infty}^{\infty}f(y) g(z - y) dy\:.
\end{align}
Therefore, by the convolution formula in \eqref{convolution}, we have
\begin{equation}\label{FMdensity}
f_{y_{t}}(z) = \int\limits_{-\infty}^{\infty} \frac{b^{N_{t}} |y|^{N_{t}-1}}{\sqrt{8 \pi \sigma_{S_{t}}^{2}}\Gamma(N_{t})} e^{-\frac{1}{2} \left( \frac{z-y-\mu_{S_{t}}}{\sigma_{S_{t}}} \right)^{2} -b |y|} dy
\end{equation}
Although not very useful in general, the expression in eq. \eqref{FMdensity} can be computationally handled with  little difficulty, a crucial fact for the concrete case study we will consider in Section \ref{CaseStudy}. In the next section we consider the $\alpha$-stable distribution model.

\section{$\boldsymbol \alpha$-Stable Distribution Model}\label{AlfaStableModel}
In Section \ref{JumpDiffusionModel}, we pointed out that the Gaussian distribution is not adequate  to model  financial data, mainly because of its slim tails, which we offset by adding jumps. In what follows, we will consider a different approach, namely we will model the data using a distribution that has fatter tails than the Gaussian one, but still preserves its most important characteristics.

\subsection{$\alpha$-Stable Distribution}
There are multiple equivalent ways to define a stable distribution. We will consider the two most common ones,  the interested reader can refer to, e.g.,  \cite{SamTaq}, for the others.
\begin{definition}
	A random variable $X$ is said to have a stable distribution if, for every $A$ and $B$ positive, there exists a positive number $C$ and a real number $D$ such that
\end{definition}
\begin{equation}
AX_{1}+BX_{2} \stackrel{\text{ d}}{=} CX+D\;,
\end{equation}
where $X_{1}$ and $X_{2}$ are independent copies of $X$ and $\stackrel{\text{ d}}{=}$ stands for {\it equal in distribution}. This implies that the sum of two stable independent identically distributed random variables is still a stable random variable, with the same distribution, up to a {\it scale factor} $C$, and a shift component $D$. As an example, we can consider  two Gaussian random variables $X_{1}$ and $X_{2}$, assumed to be independent copies of $ X \sim \mathcal{N}(\mu, \sigma^{2})$. Then, $X_{1}+X_{2} \sim \mathcal{N}(2\mu, 2\sigma^{2})$, which means that $X_{1}+X_{2} \stackrel{\text{ d}}{=} \sqrt{2}X + (2-\sqrt{2})\mu$.
Alternatively, we can define the stable distribution using characteristic functions, namely

\begin{definition}
	A random variable $X$ is said to have a stable distribution if there exist parameters $0 < \alpha \leq 2,\ \sigma \geq 0,\ |\beta| \leq 1\ \text{and}\ \mu \in \mathbb{R}$, such that its characteristic function has the form
	\begin{equation}
	\mathbb{E}(e^{i\theta X}) =
	\begin{cases}
	e^{\left[ - \sigma^{ \alpha } | \theta |^{\alpha} (1 - i \beta \sgn(\theta) \tan(\frac{\pi \alpha}{2})) + i \mu \theta \right]}, &\alpha \neq 1 \\
	e^{\left[ - \sigma | \theta | (1 + i \beta \frac{2}{\pi} \sgn(\theta) \ln |\theta| ) + i \mu \theta \right]}, &\alpha = 1
	\end{cases}
	\:.
	\end{equation}
\end{definition}
We call $\alpha$ the \emph{stability} parameter, $\beta$ the \emph{skewness} parameter, $\gamma$ the \emph{scale} parameter and $\mu$ the \emph{location} parameter. For $\alpha=2$ we obtain the normal distribution, which is the only member of the stable distribution family that has finite variance. For $\alpha \in (1,2)$ we have infinite variance and mean $\mu$, while, for $\alpha \in (0,1]$, both the mean and the variance are undefined. We note that in general there is not a solution in closed form  for the probability density function of a stable distribution. The stable distribution will be denoted by $\mathcal{S}_{\alpha,\beta}(\gamma,\mu)$ for the remainder of the paper. 

\begin{figure}[!h]
	\includegraphics[width = \columnwidth, height = .25\paperheight]{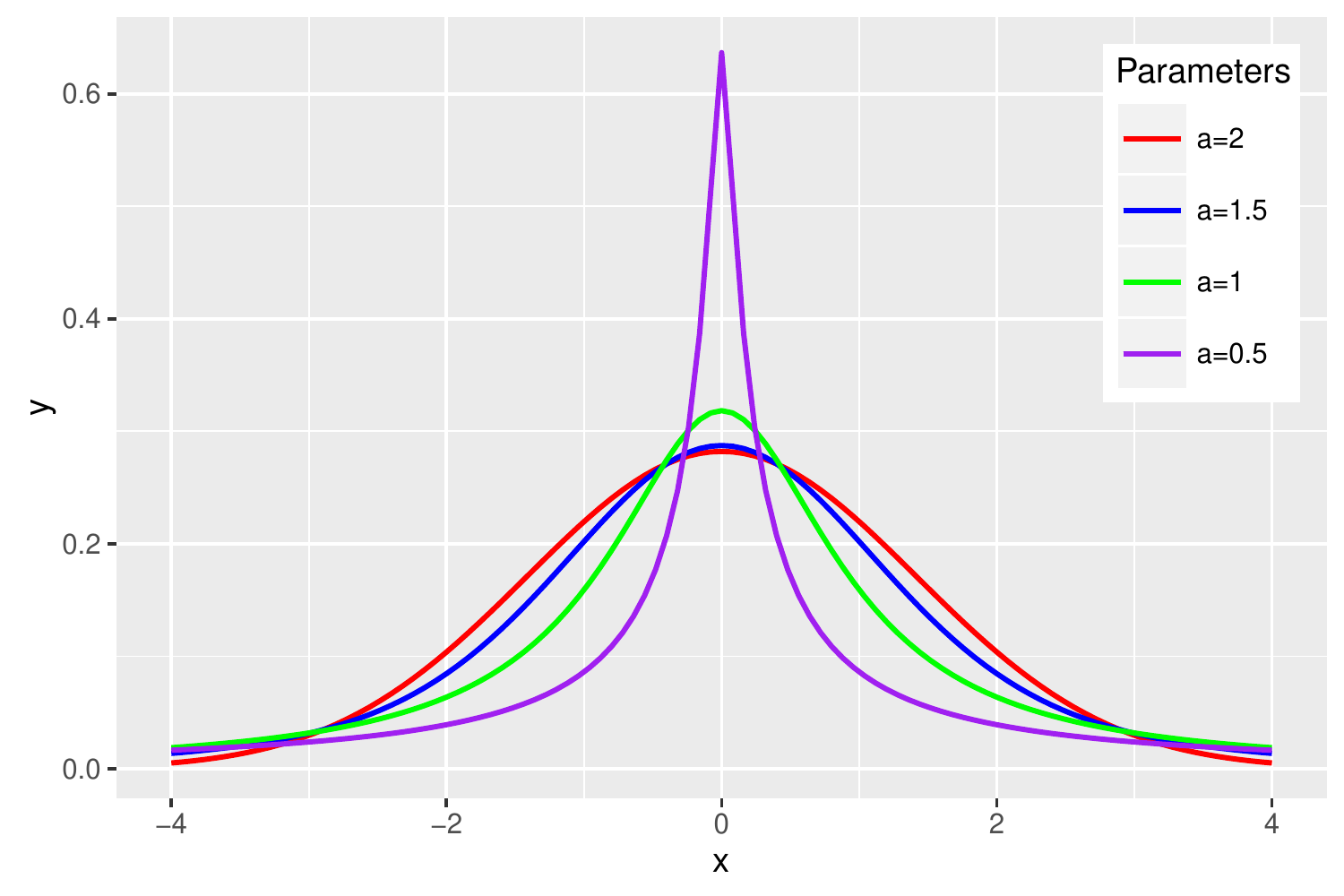}
	\caption{ \small{Comparison of the alpha-stable distribution for different values of $\alpha$. In the special case $\alpha=2$ we have a normal distribution.} }
\end{figure}

\subsection{The Model}
In the model we propose, the data are assumed to follow a symmetric $\alpha$-stable distribution, more precisely $y_{t} \sim\mathcal{S}_{\alpha,0}(\gamma_{S_{t}},\mu_{S_{t}})$. The full model is presented in the following
\begin{equation}\label{alfamodel}
\begin{cases}
y_{t} \sim\mathcal{S}_{\alpha,0}(\gamma_{S_{t}},\mu_{S_{t}}) \\
\gamma_{S_{t}} = \gamma_{j}\ \text{if}\ S_{t}=j,\ \forall j \in \{1,...,M\} \\
\mu_{S_{t}} = \mu_{j}\ \text{if}\ S_{t}=j,\ \forall j \in \{1,...,M\} \\
\alpha \in (1,2) \\
S_{t} \in \{1,...,M\} \\
p_{ij} = \prob (S_{t}=j|S_{t-1}=i) \\
\pi_{0} = [\prob (S_{0}=1),...,\prob (S_{0}=M)]
\end{cases}
\:.
\end{equation}
The motivation behind the choice of the model represented by \eqref{alfamodel}, mainly relies on empirical observations of financial data which exhibit fat tails that  can not be well described using the Gaussian approach. In particular we believe that such phenomenon can be suitably addressed by exploiting $\alpha$-stable distributions with $\alpha \in (1,2)$. Moreover, financial data often exhibit structural breaks because of abrupt changes in the market, e.g. as during the sub-prime mortgage credit crisis of 2008, which is the reason why  we consider both the scale and the location parameters, to be state-dependent.
As we  previously mentioned, in general there is no closed form for the density of an $\alpha$-stable distribution. Nevertheless, this problem can be circumvented using the fact that $y_{t}$ can be conditionally represented as a Gaussian random variable, see, e.g.,  \cite{GoKuRu10,SamTaq}, by  introducing a random variable $\lambda$ and using the property
\begin{align}
\label{conditionalnormal}
\begin{split}
\text{If:}\ \qquad &\lambda \sim \mathcal{S}_{\tfrac{\alpha}{2},1}(2 \left( \cos(\frac{\pi \alpha}{4}) \right)^{\frac{2}{\alpha}} ,0) \\
\text{then:}\ \qquad &y_{t}|\lambda \sim \mathcal{N}(\mu_{S_{t}}, \lambda\gamma^{2}_{S_{t}})\;,
\end{split}
\end{align}
which allows us to have an analytic likelihood function which significantly speeds up the sampling process. 

\begin{figure}[!h]
	\includegraphics[width = \columnwidth, height = .3\paperheight]{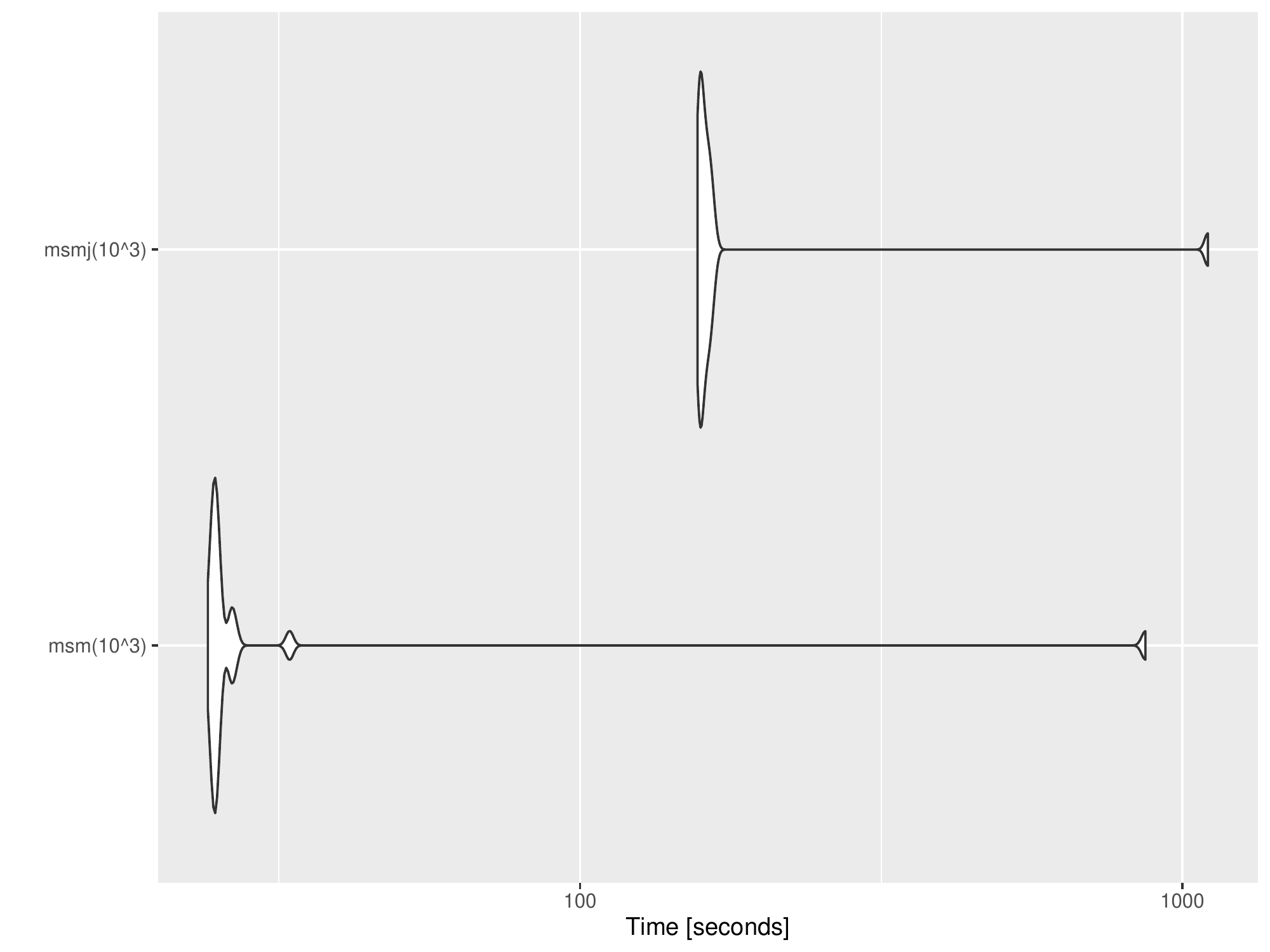}
	\caption{ \small{Comparison between the time needed for the jump diffusion model (top) and the stable distribution model (bottom) to draw 1000 samples. The best performance of the first model was $156$ seconds, its worst performance $1103$ seconds, with a mean of $207$ and a median of $160$. The  best performance of the second model was $24$ seconds, its worst performance $868$ seconds, with a mean of $68$ and a median of $25$ seconds.} }
	\label{fig:benchmark}
\end{figure}

Analogously to what we  considered in Section \ref{JumpDiffusionModel}, we have one mean for each state, without further restrictions, namely
\begin{equation}
\mu_{S_{t}} = \mu_{j}\ \text{if}\ S_{t}=j,\ \forall j \in \{1,...,M\}\;.
\end{equation}
We also want the scale parameter to be increasing with respect to the state, namely
\begin{equation}
\gamma_{S_{t}} =
\begin{cases}
\gamma_{1} , &S_{t} = 1 \\
\gamma_{1} \prod\limits_{i=2}^{S_{t}} (1+h_{i}), &S_{t}=j,\ 2 \leq j \leq M
\end{cases}
\;,
\end{equation}
where $\forall i \in \{2,3,...,M\},\ h_{i}>0$, which leads to the property
\begin{equation}
\gamma_{1} < \gamma_{2} < ... < \gamma_{M}\;,
\end{equation}
so that  an increase in the state number indicates an increase in volatility.

\section{Implementation}
In this section we get into the specifics of our two models. In particular we provide the details regarding the likelihood functions, the priors and the posteriors, for both the {\it Jump Diffusion Model}, described in Section \ref{JumpDiffusionModel}, and for the {\it $\alpha-$Stable Model}, defined in Section \ref{AlfaStableModel}.
The concept of duration analysis is also explained, along with its importance. Both of the models will be characterized by four states, with the states being interpreted as \emph{low, medium, high} and \emph{very high} volatility {\it regime}. Let us start by defining the following quantities
\begin{align}
y^{j} & = \{y_{t} \in \tilde{y}_{T} :S_{t}=j\},\ j \in \{1,2,3,4\} \;,\\
n_{j} & = \# y^{j}\:.
\end{align}
For the rest of this section we will suppress unneeded parameters. Therefore, e.g., the conditional posterior $f(\mu_{j}|\boldsymbol y, \sigma_{j}, N_{j},...)$, will be denoted by $f(\mu_{j}|\boldsymbol y)$, the general rule being that the parameters that are not being inferred on are considered known.
\subsection{Jump Diffusion Model}
\label{subsec:jump_diffusion_model}
The description  of the implementation is divided into three parts, namely: the first part deals with the form of the likelihood function, the second with the priors while the third part provides a detailed analysis of the different types of obtained posteriors. 
\subsubsection{Likelihood}
\label{likelihood1}
We have to take into account whether there are jumps in the model or not, as well as the state of each observation. Hence, if $N_{j} = 0$, we define
\begin{equation}
\begin{aligned}
L^{j}\left( \cdot \right) &= \prod\limits_{y_{t} \in y^{j}} \frac{e^{- \frac{1}{2} \left( \frac{y_{t}-\mu_{j}}{\sigma_{j}} \right)^{2}} }{\sqrt{2 \pi \sigma_{j}^{2}}} \\
& = \left( 2 \pi \sigma_{j}^{2} \right)^{-\frac{n_{j}}{2}} e^{-\frac{1}{2 \sigma_{j}^{2}} \sum\limits_{y_{t} \in y^{j}}\left( y_{t}-\mu_{j} \right)^{2} }\;,
\end{aligned}
\end{equation}
while, if $N_{j} \geq 1$, we define $L^{j}\left( \cdot \right)$ as
\begin{equation}
\prod\limits_{y_{t} \in y^{j}} \frac{b^{N_{j}}}{\sqrt{8 \pi \sigma_{j}^{2}}\Gamma(N_{j})} \int\limits_{-\infty}^{\infty} |y|^{N_{j}-1}e^{-\frac{1}{2}\left( \frac{y_{t}-y-\mu_{j}}{\sigma_{j}} \right)^{2}}dy\:.
\end{equation}
Then, the full likelihood function is 
\begin{equation}
L_{n} = \prod\limits_{j=1}^{4} L^{j}\:,
\end{equation}
which has a standard form only if $N_{j}=0$, for every $j \in \{1,2,3,4\}$. As this very rarely happens, we will use the Metropolis-Hastings algorithm in this model.

\subsubsection{Priors}
\begin{raggedleft}
	\textbf{Mean:}
\end{raggedleft} 
We take the mean to be normally distributed. Moreover, we give the same prior to the means of all the states, namely
\begin{equation}
\label{meanprior}
\mu_{j} \sim \mathcal{N}\left(0, \frac{1}{k}\right),\ 1 \leq j \leq 4 \ \text{and}\ k>0\:.
\end{equation}

\begin{raggedleft}
	\textbf{Variance:}
\end{raggedleft} 
The variance $\sigma_{1}^{2}$ will have an inverse-gamma prior.
\begin{equation*}
\sigma_{1}^{2} \sim\ \text{inv}\Gamma(\alpha_{0}, \beta_{0}),\ \alpha_{0}, \beta_{0}>0\:.
\end{equation*}

\begin{raggedleft}
	\textbf{H parameters:}
\end{raggedleft}
We previously saw that in order for \eqref{sigmainequality} to hold, we need $h_{j} > 0,\ \forall j \in \{2,3,4\}$, hence we define $h_{j}^{\star} := 1+h_{j}$, for all $j$, and make these parameters Fr\'{e}chet distributed, namely 
\begin{equation}
h_{j}^{\star} \sim \mathcal{F}(h_{j}^{\star}|1, \alpha_{F}, s_{F})\;,
\end{equation}
then the density function of $h_{j}^{\star}$ reads as follows
\begin{equation}
f(h_{j}^{\star}) = \frac{\alpha_{F}}{s_{F}} \left( \frac{h_{j}^{\star}-1}{s_{F}} \right)^{1-\alpha_{F}} e^{- \left( \frac{h_{j}^{\star}-1}{s_{F}} \right)^{- \alpha_{F}}}\;,
\end{equation}
where $\alpha_{F},s_{F}>0$, and $f$ is defined for $h_{j}^{\star}>1$.

\begin{raggedleft}
	\textbf{Poisson parameters:}
\end{raggedleft} 
For the  priors of the Poisson parameters we refer to \eqref{poissonprior}.

\begin{raggedleft}
	\textbf{Transition probabilities:}
\end{raggedleft}
For the transition probabilities we will use a Dirichlet prior, namely 
\begin{equation}
(p_{1j},p_{2j},p_{3j},p_{4j}) \sim \mathcal{D}(m_{1j},m_{2j},m_{3j},m_{4j})\;,
\end{equation}
for every $j \in \{1,2,3,4\}$. The density function of this particular Dirichlet distribution is given by
\begin{align}
f(p_{1j},p_{2j},p_{3j},p_{4j}) = \frac{\Gamma\left(\sum\limits_{i=1}^{4} m_{ij} \right)}{\prod\limits_{i=1}^{4}\Gamma(m_{ij})} \prod\limits_{i=1}^{4} p_{ij}^{m_{ij}-1}
\;,
\end{align}
and it is defined on the simplex
\begin{align}
\nonumber & p_{1j},p_{2j},p_{3j} > 0 \;, \\
& p_{1j}+p_{2j}+p_{3j} < 1 \;,\\
\nonumber & p_{4j} = 1 - p_{1j}+p_{2j}+p_{3j}
\;,
\end{align}
while, everywhere else, its value is zero. Finally, the parameter $b$ is a constant.

\subsubsection{Posteriors}
\begin{raggedleft}
	\textbf{Mean:}
\end{raggedleft}
Because the likelihood function depends on whether or not jumps have occurred, we have two different posteriors for the mean. In particular, if $N_{j}=0$, by \eqref{nn}, we have
\begin{equation}
\label{meanposterior}
\mu_{j}|y^{j} \sim \mathcal{N} \left(\frac{n_{j}\bar{y}^{j}}{n_{j}+k\sigma_{j}^{2}},\frac{\sigma_{j}^{2}}{n_{j}+k \sigma_{j}^{2}} \right)\;,
\end{equation}
for all $j \in \{1,2,3,4\}$, while, if $N_{j} \geq 1$, we obtain
\begin{equation}
f(\mu_{j}|y^{j}) \propto f(\mu_{j}) L^{j}(\mu_{j}) \:.
\end{equation}

\begin{raggedleft}
	\textbf{Variance:}
\end{raggedleft}
Similarly to the previous point, we have to differentiate between the jump and no-jump cases. Therefore, if $N_{j}=0$, by \eqref{ign}, we have
\begin{equation}
\sigma_{1}^{2} \sim\ \text{inv}\Gamma \left(\frac{n_{1}}{2}+\alpha_{0}, \frac{1}{2} \sum\limits_{y_{t} \in y^{1}} \left( y_{t}-\mu_{1} \right)^{2} + \beta_{0} \right)\;,
\end{equation}
otherwise, we obtain
\begin{equation}
f(\sigma_{1}|y^{1}) \propto f(\sigma_{1}) L^{1}(\sigma_{1})\:.
\end{equation}

\begin{raggedleft}
	\textbf{H parameters:}
\end{raggedleft}
In order to obtain $h_{i},\ i \in \{2,3,4\}$, we need to transform the data, also taking into account the  different states. 

In particular we have the following cases

\begin{raggedleft}
	$\boldsymbol{S_{t}=2:}$
\end{raggedleft}
\begin{align*}
y_{t} \sim \mathcal{N}(\mu_{2}, \sigma_{1}^{2} h_{2}^{\star}) \Rightarrow  \psi_{t} := \frac{y_{t}-\mu_{2}}{\sigma_{1}} \sim \mathcal{N}(0, h_{2}^{\star})\:,
\end{align*} 
where we have used the transformed data set $\psi_{t}$  to obtain a posterior for $h_{2}^{\star}$ when $N_{2}=0$, namely
\begin{equation}
f(h_{2}^{\star}|y^{2}) \propto \left( h_{2}^{\star} \right)^{- \frac{n_{2}}{2}} e^{- \frac{1}{2 h_{2}^{\star}} \sum\limits_{i=1}^{n_{2}} \psi^{2}_{i} } \mathcal{F}(h_{2}^{\star}|1,\alpha_{F},s_{F})\:.
\end{equation}

\begin{raggedleft}
	$\boldsymbol{S_{t}=3:}$
\end{raggedleft}
\begin{equation*}
y_{t}  \sim \mathcal{N}(\mu_{3}, \sigma_{1}^{2} h_{2}^{\star} h_{3}^{\star}) \Rightarrow  \zeta_{t} := \frac{y_{t}-\mu_{3}}{\sqrt{\sigma_{1}^{2} h_{2}^{\star}}} \sim \mathcal{N}(0, h_{3}^{\star})\;,
\end{equation*} 
which gives us the posterior
\begin{equation}
f(h_{3}^{\star}|y^{3}) \propto \left( h_{3}^{\star} \right)^{- \frac{n_{3}}{2}} e^{- \frac{1}{2 h_{3}^{\star}} \sum\limits_{i=1}^{n_{3}} \zeta^{2}_{i} } \mathcal{F}(h_{3}^{\star}|1,\alpha_{F},s_{F})\;,
\end{equation}
hence, when $N_{3}=0$, the posterior is analogous to the one for $h_{2}^{\star}$, with the only difference being that we use $\zeta_{t}$ instead of $\psi_{t}$. 
\begin{flushleft}
	$\boldsymbol{S_{t}=4:}$
\end{flushleft}
\begin{align*}
y_{t} &\sim \mathcal{N}(\mu_{4}, \sigma_{1}^{2} h_{2}^{\star} h_{3}^{\star} h_{4}^{\star})\\ \Rightarrow  \theta_{t} &:= \frac{y_{t}-\mu_{4}}{\sqrt{\sigma_{1}^{2} h_{2}^{\star} h_{3}^{\star}}} \sim \mathcal{N}(0, h_{4}^{\star})\;,
\end{align*} 
which yields the posterior
\begin{equation}
f(h_{4}^{\star}|y^{4}) \propto \left( h_{4}^{\star} \right)^{- \frac{n_{4}}{2}} e^{- \frac{1}{2 h_{4}^{\star}} \sum\limits_{i=1}^{n_{4}} \theta^{2}_{i} } \mathcal{F}(h_{4}^{\star}|1,\alpha_{F},s_{F})\:.
\end{equation}

In the case where we have jumps, i.e. $N_{j} \geq 1,\ j \in \{2,3,4\}$, there is no analytic expression, therefore 
\begin{equation*}
f(h_{j}^{\star}|y^{j}) \propto \mathcal{F}(h_{j}^{\star}|1,\alpha_{F},s_{F}) L^{j}(h_{j}^{\star})\:.
\end{equation*}

\begin{raggedleft}
	\textbf{Poisson parameters:}
\end{raggedleft}
Concerning the posterior of the theta parameters, for $j \in \{1,2,3,4\}$, we have
\begin{equation}
\begin{aligned}
f(\theta_{j}|y^{j}) \propto f(y^{j}|\theta_{j}) = \sum_{N_{j}=0}^{\infty} f(y^{j},N_{j}|\theta_{j}) \\ \propto \sum_{N_{j}=0}^{\infty} f(N_{j}|\theta_{j})f(y^{j}|N_{j},\theta_{j}) = \sum_{N_{j}=0}^{\infty} f(N_{j}|y^{j},\theta_{j})
\end{aligned}
\end{equation}

\begin{raggedleft}
	\textbf{Transition probabilities:}
\end{raggedleft}
The transition probabilities differ from the other parameters in that they do not depend directly on the observations $y_{1},...,y_{T}$. Instead, they depend on the vector of states $\tilde{S}_{T}$. Assuming that the vector $\tilde{S}_{T}$ is known, the posterior distribution of the transition probability vector $(p_{1j},p_{2j},p_{3j},p_{4j})$, $j \in \{1,2,3,4\}$, has the Dirichlet distribution
\begin{equation}
\mathcal{D}(m_{1j}+n_{1j},m_{2j}+n_{2j},m_{3j}+n_{3j},m_{4j}n_{4j})\;,
\end{equation}
where $n_{ij}$ is the number of transitions from state $j$ to state $i$.

\subsection{$\alpha$-Stable Distribution Model}
In what follows, we proceed analogously to subsection \ref{subsec:jump_diffusion_model}.
\subsubsection{Likelihood}
Using the fact that, in the present setting, our data are conditionally normal, see \eqref{conditionalnormal},  the likelihood function reads as follows
\begin{align*}
L \left( \cdot|\boldsymbol y \right) &= \prod\limits_{j=1}^{4} \prod\limits_{y_{t} \in y^{j}} \frac{e^{- \frac{1}{2} \left( \frac{y_{t}-\mu_{j}}{\lambda\gamma_{j}}\right)^{2}} }{\sqrt{2 \pi \lambda \gamma_{j}^{2}}} \\
& = \prod\limits_{j=1}^{4} \left( 2 \pi \lambda \gamma_{j}^{2} \right)^{-\frac{n_{j}}{2}} e^{-\frac{1}{2 \lambda \gamma_{j}^{2}} \sum\limits_{y_{t} \in y^{j}}\left( y_{t}-\mu_{j} \right)^{2} }\;,
\end{align*}
and, unlike in the previous model, we do not have to worry about multiple cases.

\subsubsection{Priors}
\begin{raggedleft}
	\textbf{Mean:}
\end{raggedleft}
The prior of the means is the same as in eq.~\eqref{meanprior}.

\begin{raggedleft}
	\textbf{Scale:}
\end{raggedleft}
The distribution of the scale is analogous to that of the variance in the previous model, namely
\begin{align*}
\gamma^{2} \sim\ \text{inv}\Gamma(\alpha_{1}, \beta_{1}), \alpha_{1}, \beta_{1}>0\:.
\end{align*}

\begin{raggedleft}
	\textbf{H parameters:}
\end{raggedleft}
These parameters are exactly the same as they were in the previous model, in fact their role remains unchanged, since they allow for the volatility to increase as the states increase. 

\begin{raggedleft}
	\textbf{Lambda:}
\end{raggedleft}
The lambda parameter follows a stable distribution, hence
\begin{align*}
\lambda \sim S_{\tfrac{\alpha}{2},1}(2 \left( \cos\left(\frac{\pi \alpha}{4}\right) \right)^{\frac{2}{\alpha}} ,0)\:.
\end{align*}

\subsubsection{Posteriors}
\begin{raggedleft}
	\textbf{Mean:}
\end{raggedleft}
The posterior of the mean is analogous to that of the one in eq. \eqref{meanposterior}, with the only difference being the form of the variance. In particular, we have
\begin{equation}
\mu_{j}|y^{j} \sim \mathcal{N} \left(\frac{n_{j}\bar{y}^{j}}{n_{j}+k \lambda \gamma_{j}^{2}},\frac{\lambda \gamma_{j}^{2}}{n_{j}+k \lambda \gamma_{j}^{2}} \right)
\:.
\end{equation}

\begin{raggedleft}
	\textbf{Scale:}
\end{raggedleft}
The posterior of the scale is 
\begin{equation}
\gamma^{2} \sim\ \text{inv}\Gamma \left(\frac{n_{1}}{2}+\alpha_{1}, \frac{1}{2} \sum\limits_{y_{t} \in y^{1}} \left( y_{t}-\mu_{1} \right)^{2} + \beta_{1} \right)
\:.
\end{equation}

\begin{raggedleft}
	\textbf{H parameters:}
\end{raggedleft}
In what follows we limit ourselves to listing the needed transformations, therefore we have

\begin{raggedleft}
	$\boldsymbol{S_{t}=2:}$
\end{raggedleft}
\begin{align*}
y_{t} \sim \mathcal{N}(\mu_{2},\lambda \gamma^{2} h_{2}^{\star}) \Rightarrow  \psi_{t} := \frac{y_{t}-\mu_{2}}{\sqrt{\lambda \gamma^{2}}} \sim \mathcal{N}(0, h_{2}^{\star})\:.
\end{align*} 

\begin{raggedleft}
	$\boldsymbol{S_{t}=3:}$
\end{raggedleft}
\begin{equation*}
y_{t}  \sim \mathcal{N}(\mu_{3}, \lambda \gamma^{2} h_{2}^{\star} h_{3}^{\star}) \Rightarrow  \zeta_{t} := \frac{y_{t}-\mu_{3}}{\sqrt{\lambda \gamma^{2} h_{2}^{\star}}} \sim \mathcal{N}(0, h_{3}^{\star})\:.
\end{equation*} 

\begin{raggedleft}
	$\boldsymbol{S_{t}=4:}$
\end{raggedleft}
\begin{align*}
y_{t} &\sim \mathcal{N}(\mu_{4}, \lambda \gamma^{2} h_{2}^{\star} h_{3}^{\star} h_{4}^{\star}) \\
\Rightarrow  \theta_{t} &:= \frac{y_{t}-\mu_{4}}{\sqrt{\lambda \gamma^{2} h_{2}^{\star} h_{3}^{\star}}} \sim \mathcal{N}(0, h_{4}^{\star})\:.
\end{align*} 
The posteriors are obtained as in the previous case.

\begin{raggedleft}
	\textbf{Lambda:}
\end{raggedleft}
Since there is no closed form for the posterior distribution of the lambda parameter,  we only write
\begin{equation}
\label{eq:lambda_posterior}
f(\lambda|\boldsymbol y) \propto S_{\tfrac{\alpha}{2},1}(2 \left( \cos\left(\frac{\pi \alpha}{4}\right) \right)^{\frac{2}{\alpha}} ,0) L(\lambda|\boldsymbol y)
\:.
\end{equation}

The prior and posterior of the transition probabilities are the same in both proposed models. In fact, the transition probabilities do not depend by any of the parameters, but only by the state vector $S_{T}$.

\subsection{Duration Analysis}
\label{subsubsec:duration}
The expected duration of each state for a MSM is a quantity of significant interest. Having an estimate of how long a certain data set remains in a particular state can give us useful insights into how the model will behave for a certain period of time. In this section we are going to explain how the expected duration can be  calculated exploiting the transition probabilities.

The expected duration, denoted by $\hat{d}_j$, is  defined as follows
\begin{equation}
\label{eq:expdur_def}
\hat{d}_j := \expect\left[d_j\right]\;,
\end{equation}
where $d_j$ is the random variable that models the length of the time interval for which the time series is in state $j$. The first thing we have to consider is $\prob\left(d_j = d\right)$, the probability of the data being in state $j$, meaning
\begin{equation}
\nonumber
\prob\left(d_j = d\right) = p_{jj}^{d-1} \left(1-p_{jj}\right)\;,
\end{equation}
where $p_{ij} = \prob\left(S_t = j| S_{t-1} = i\right)$. It just so happens that $\hat{d}_j$ has a very simple closed form, in particular
\begin{align}
\nonumber\expect\left(d_j\right) &= \sum_{m=1}^\infty m p_{jj}^{m-1} \left(1-p_{jj}\right) \\
\nonumber&= \left(1-p_{jj}\right) \lim_{n\to\infty} \sum_{m=1}^{n} m\, p_{jj}^{m-1} \\
\nonumber&= \left(1-p_{jj}\right) \lim_{n\to\infty} \frac{1}{p_{jj}} \sum_{m=1}^{\infty} m\, p_{jj}^m \\
\nonumber&= \left(1-p_{jj}\right) \lim_{n\to\infty} \frac{1-(n+1)p_{jj}^n+n p_{jj}^{n+1}}{(1-p_{jj})^2}\\
\nonumber&= \left(1-p_{jj}\right) \frac{1}{(1-p_{jj})^2}\\
\label{eq:expdur}&= \frac{1}{1-p_{jj}}.
\end{align}

We will use the previous expression later on, when we  compare the state durations obtained in the present paper, with  those provided in \cite{DiPMF16}. We will see that there is a significant difference in the state durations, showing that  the models developed in this paper perform better than the one proposed in \cite{DiPMF16}  to model the time series of the Chicago Board Options Exchange Volatility Index, better known as VIX.

\section{Case Study}\label{CaseStudy}
Our case study is concerned with the application of the above theory to developing an indicator that has a role similar to the one played by the VIX. In particular we use the set of S\&P500 weekly prices, considering a time interval that runs from the 3rd of January, 2007 to 29th of December, 2014. We picked this interval to include the sub-prime mortgage crash of 2008 as well as the subsequent period of relative calm. This choice allows us to analyse how our approach  performs in both situations. We will show that our techniques  improve the results stated in \cite{DiPMF16}, where the model was very effective in periods of high volatility, but also too smooth in case of low volatility. Our results are summarised below with respect to both the {\it Jump Diffusion Model}, defined in Section \ref{JumpDiffusionModel}, and the 
{\it $\alpha-$Stable Model}, provided in Section \ref{AlfaStableModel}.

\subsection{Jump Diffusion Model}
\label{jump_diffusion_case}
\begin{figure}[!h]
	\includegraphics[width = \columnwidth, height = .35\paperheight]{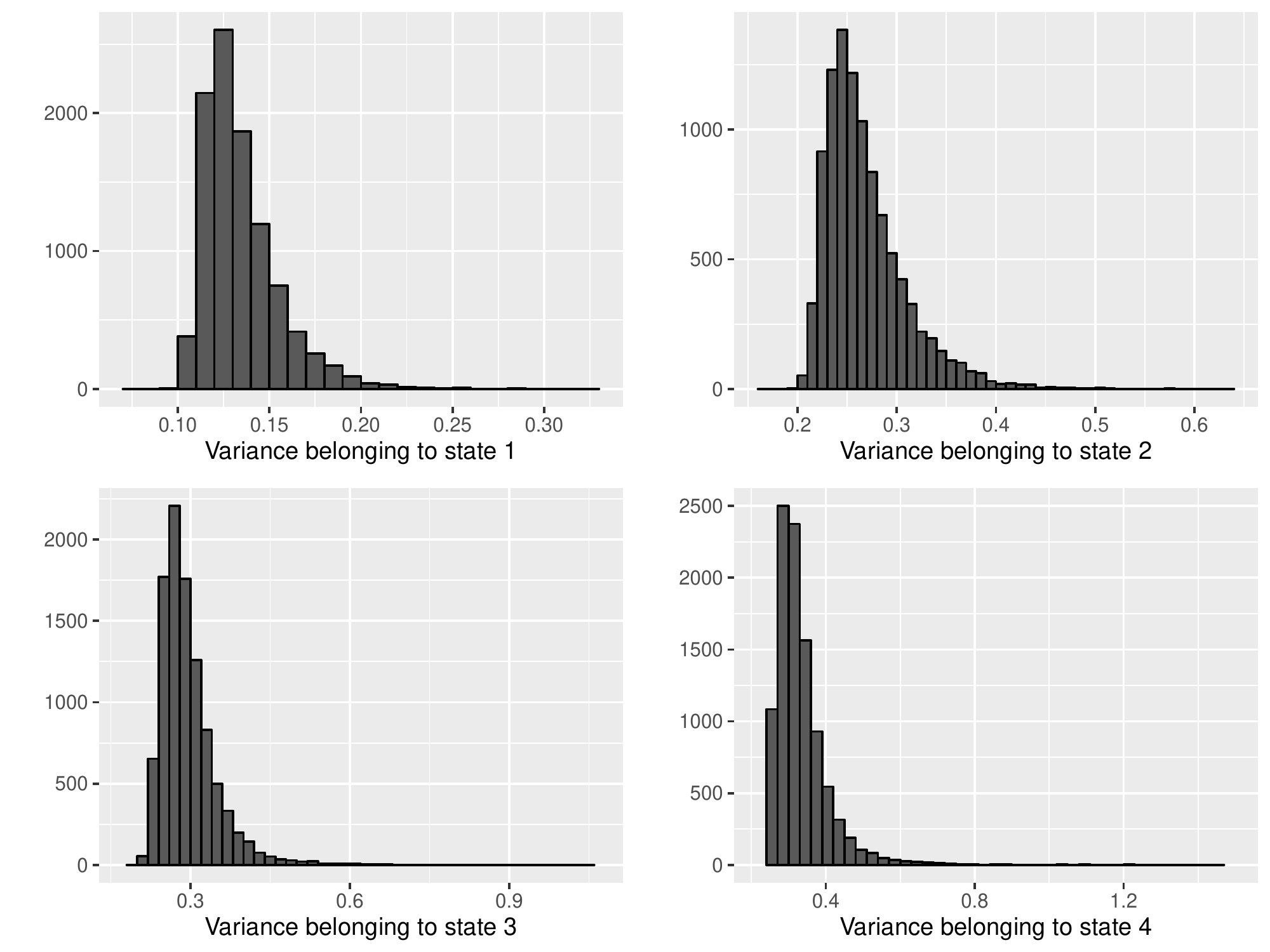}
	\caption{ \small{Comparison between the histograms of each of the variances belonging to the four states.} }
	\label{fig:variance_comparison}
\end{figure}

For the jump diffusion model we model the data as a zero mean process in order to make the framework more parsimonious. We take the exponential distribution parameter $b$ to be equal to 40, in order to make the contribution of each extra jump to the variance relatively small. This choice of b allows for a finer analysis. We first present the histograms of the sampled variances, see Fig. \ref{fig:variance_comparison}. 

As we can see, the algorithm is rather accurate in sampling the variances. In particular, we recall that the theoretical posterior of the variances is an inverse-Gamma distribution, which is exactly what we can observe in the histograms. 

Moreover, in Table \ref{tbl:variance1}, we report the point estimates of each variance value.
\begin{table}[h!]
	\caption{Gaussian Variance Point Estimates}
	\centering
	\begin{tabular}{c|c}
		\label{tbl:variance1}
		Estimator & Value \\
		\hline
		$\hat{\sigma}_{1}^{2}$ & $0.155709$ \\
		$\hat{\sigma}_{2}^{2}$ & $0.2336135$ \\
		$\hat{\sigma}_{3}^{2}$ & $0.254559$ \\
		$\hat{\sigma}_{4}^{2}$ & $0.2816716$
	\end{tabular}
	\label{tab:variance1}
\end{table}

\begin{figure}[h!]
	\includegraphics[width = \columnwidth, height = .35\paperheight]{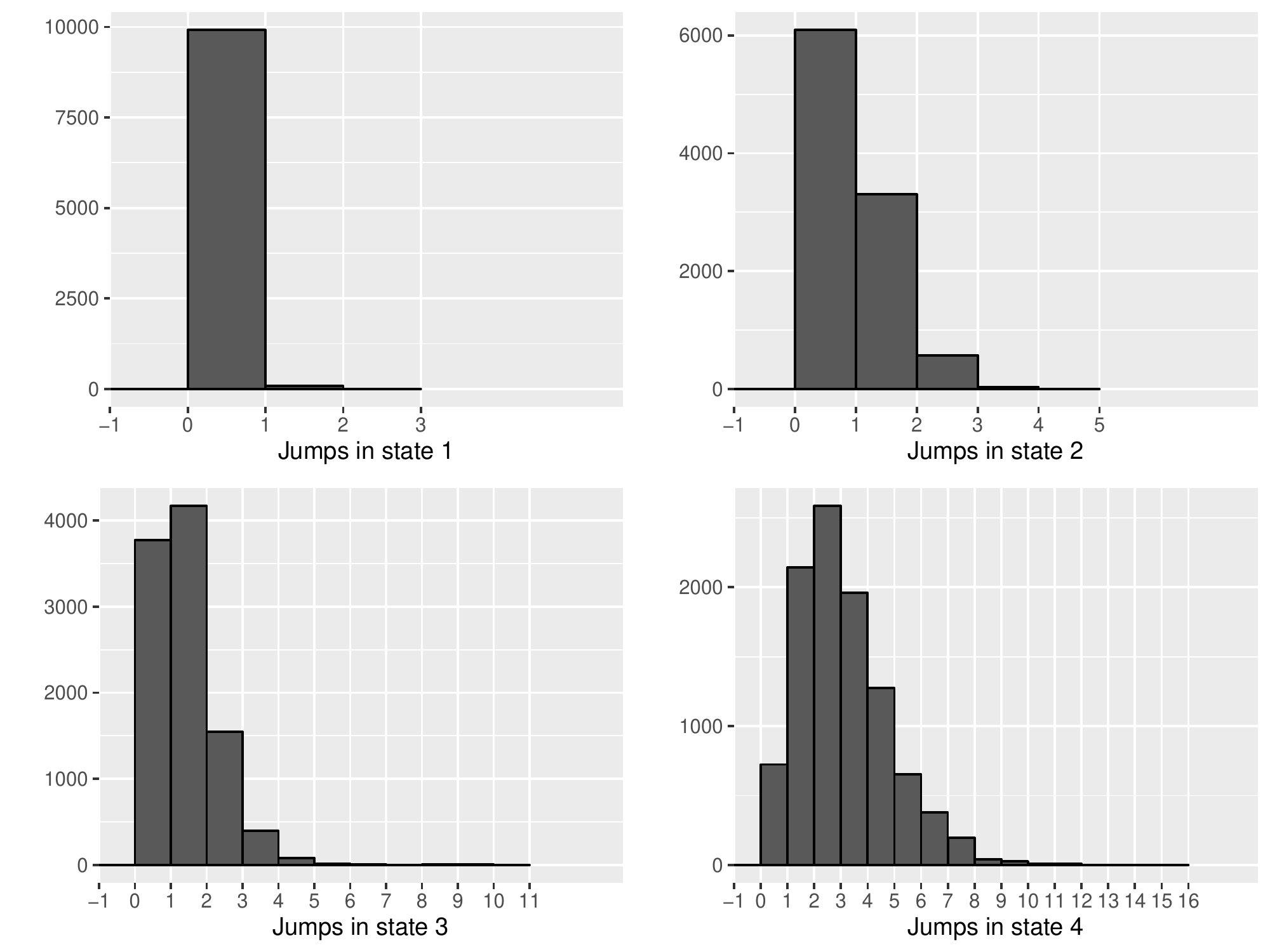}
	\caption{ \small{Comparison between the histograms of each of the variances belonging to the four states.} }
	\label{fig:jump_comparison}
\end{figure}

Concerning the distribution of the jumps, we refer to Fig. \ref{fig:jump_comparison} which highlights that the jumps, like the Gaussian variances,  they have an amplitude that increases according with the number of state. The first state almost never has jumps. The latter means that, when we are in the first state, the description of the variance of the observations  is left mainly to the parameter $\sigma_{1}^{2}$. For the other states we see an increase in non-zero jumps and also an increase in the average number of jumps, as we go up in states.

Noticing how close the sample variances $\hat{\sigma}_{2}^{2}, \hat{\sigma}_{3}^{2},\ \text{and}\ \hat{\sigma}_{4}^{2}$ are in value (Table \ref{tab:variance1}), we see that the distinction between the states is in the jumps. This is exactly the result that we have aimed  to obtain by defining this model. In \eqref{eq:transition_matrix_jump}, we report the matrix with the transition probabilities of the model, while eq. \eqref{eq:transition_matrix_jump} indicates the average number of transitions from one state to another. In particular, the element  $\hat P_{ij}$ is the transition probability $\hat p_{ij}$ and $\hat M_{ij}$ is the number of transitions from state $i$ to state $j$.

\begin{equation}
\label{eq:probability_matrix_jump}
\hat P = \begin{bmatrix}
0.6147 &  0.2522 &  0.0894 &  0.0435\\
0.5326 &  0.3073 &  0.1082 &  0.0518\\
0.1491 &  0.0805 &  0.7250 &  0.0453\\
0.1671 &  0.0904 &  0.0487 &  0.6936
\end{bmatrix}
\end{equation} 

\begin{equation}
\label{eq:transition_matrix_jump}
\hat M = \begin{bmatrix}
113.62 & 46.61 & 16.53 & 8.04 \\
46.35 & 26.74 & 9.41 & 4.51 \\
15.50 & 8.37 & 75.37 & 4.71 \\
9.38 & 5.08 & 2.74 & 38.97
\end{bmatrix}
\end{equation} 

The last thing we want to list before comparing our indicator with the VIX index, is the expected values of the state durations. Using eq. \eqref{eq:expdur}, we obtain the following result
\begin{align}
\label{duration1}
\hat{d_1} = 2.6501, \quad \hat{d_2}=1.4227\;,\\
\nonumber \hat{d_3}=3.6518, \quad \hat{d_4}=3.3432\;,
\end{align}
which are coherent with the choice we made of exploiting  the Dirichlet prior  to make the model more resistant on states 3 and 4, while allowing more transitions between states 1 and 2.

\begin{remark}
	In~\cite{DiPMF16} the authors underlined that one of the main  issues that the proposed model cannot fix was the over-smoothing effect. In particular, they obtained the following durations: $\hat{d_1}=45.2489 $, $\hat{d_2}=23.9808$ $\hat{d_3}=26.4550$ $\hat{d_4}=3.3356$. One can notice that the duration of the highest state is conserved while the others changed, which was exactly our goal. Moreover such results clearly show an important development in the solution of the aforementioned issue, since duration can be used as a quantitative indicator of smoothness.
\end{remark}

Finally, we compare our results with the VIX index data. In particular, our volatility indicator, denoted by $I_{t}^{J}$, will  indicates the expected standard deviation of the data at time $t$, namely
\begin{equation}
I_{t}^{J} := \sqrt{\sum\limits_{j=1}^{4} \prob (S_{t}=j|\psi_{t}) \left(\hat{\sigma}_{j}^{2} + \frac{\hat{N}_{j} \left( \hat{N}_{j}+1 \right)}{40^{2}} \right)} \;, 
\end{equation} 
where $J$ stands for {\it jump},  we also present a visual comparison in the figure below.

\begin{figure}[!h]
	\includegraphics[width = \columnwidth, height = .25\paperheight]{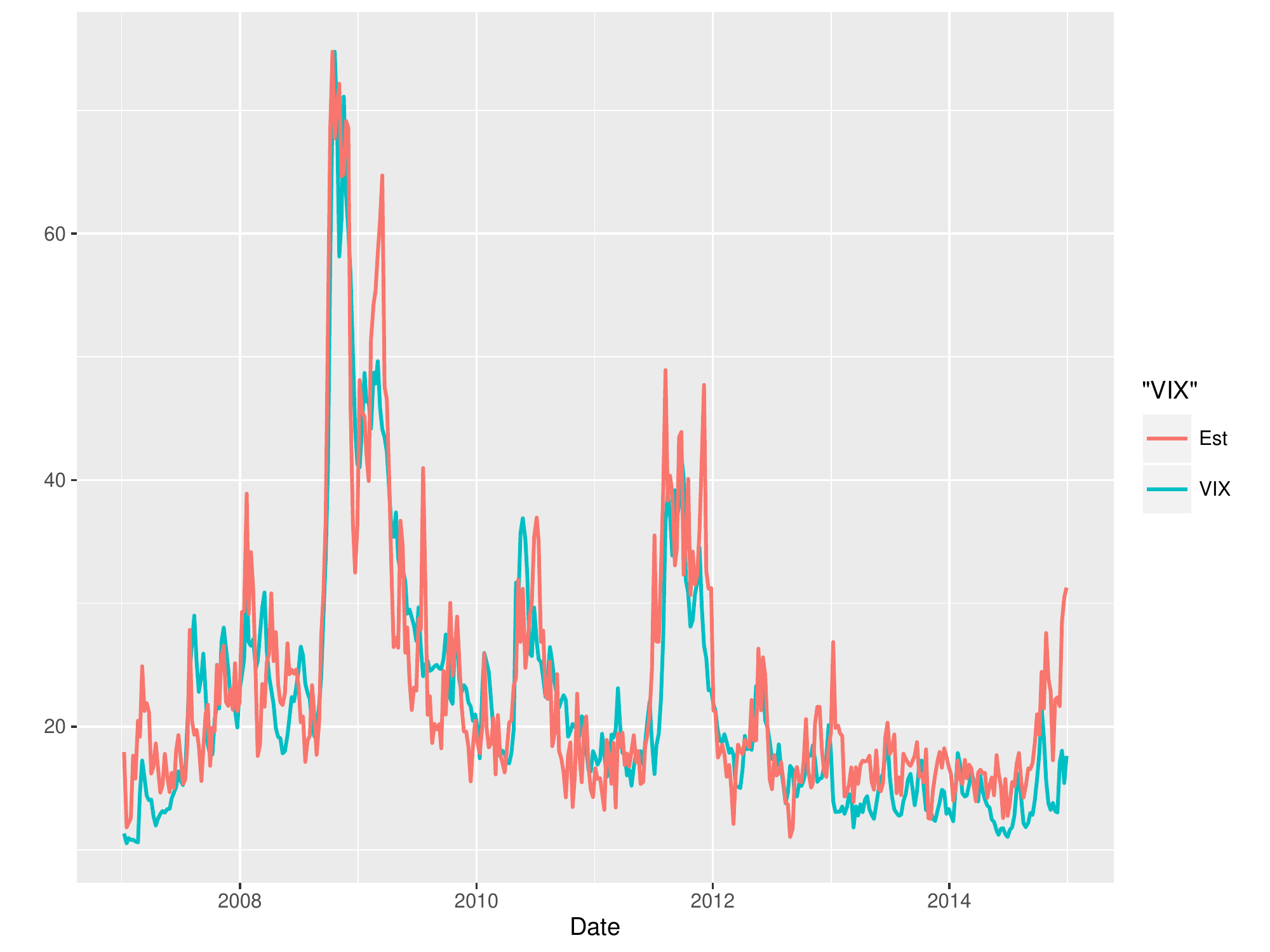}
	\caption{ \small{Visual comparison between the VIX and the volatility indicator $I_{t}^{J}$. We have applied a linear scaling function to $I_{t}^{J}$.} }
\end{figure}

Since we want to compare this model to the $\alpha-$stable distribution model that follows, we need some way to quantify what is a {\it good} estimate of the volatility. Following a standard approach, we take into consideration the sum of the squares of the difference between our volatility indicator and the VIX, namely we define 
\begin{equation}
\label{eq:squares1}
\mathbf{S}^{J} = \sum\limits_{t=1}^{T} \left( I_{t}^{J} - y_{t, \text{\tiny VIX}} \right)^{2}\;,
\end{equation}
where $y_{t, \text{\tiny VIX}}$ is the value of the VIX at time $t$. Using eq. \eqref{eq:squares1}, we obtain $\mathbf{S}^{J}=26.23$. In the next section we analyse the performances provided by  the second model.

\subsection{$\alpha$-Stable Distribution Model}
\label{alpha_stable_case}
\begin{figure}[!h]
	\includegraphics[width = \columnwidth, height = .3\paperwidth]{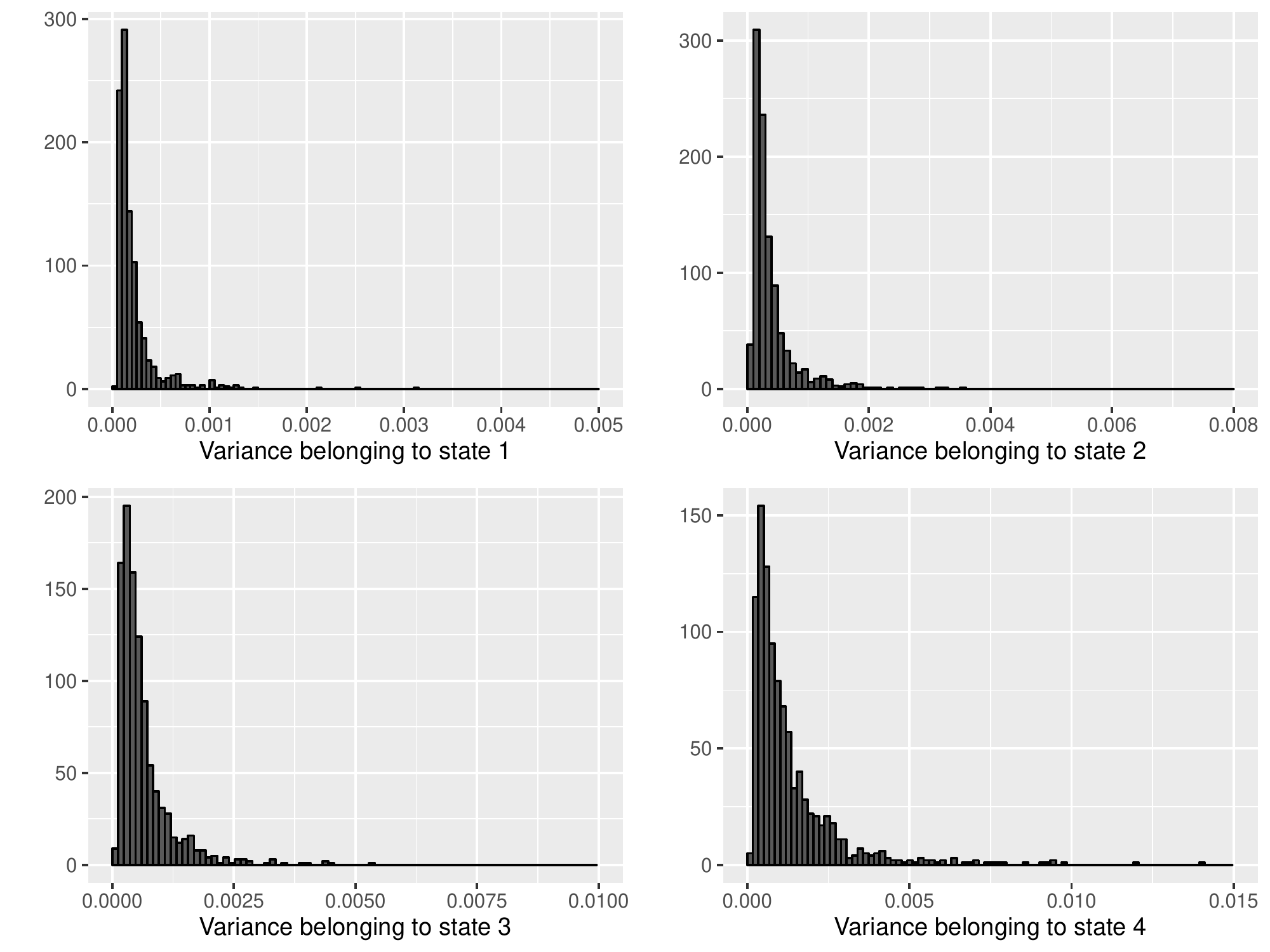}
	\caption{ \small{The histograms of the variances $\lambda \gamma_{S_{t}}^{2}$.} }
	\label{fig:stable_variance_histograms}
\end{figure}

In this model we do not have to sample the jumps $N_{t}$, or their parameters $\theta_{S_{t}}$. We use this decrease in the number of parameters  to infer on the mean $\mu_{S_{t}}$ without making the simulations too cumbersome, or inaccurate. In Fig. \ref{fig:stable_variance_histograms}, we present the histograms of the values $\lambda \gamma_{j}^{2},\ j \in \{1,2,3,4\}$. We choose these values, instead of just $\gamma_{j}^{2}$, since the indicate the variance of $y_{t}|\lambda$. Related estimates can be found in Table \ref{tbl:variance2}. 

Looking at  Table \ref{tbl:variance2}, we notice that there is a much bigger difference between the last three scale estimators, than there one between the last three Gaussian variance estimators. The latter result is due to the fact that the present model lacks of a jump component, therefore all the volatility has to be {\it explained} by mean of the scale parameters. 

\begin{table}[h!]
	\caption{Parameter Point Estimates}
	\centering
	
	\begin{tabular}{c|c}
		\label{tbl:variance2}
		Estimator & Value \\
		\hline
		$\hat{\lambda}$ & $0.00390925$ \\
		$\hat{\gamma}_{1}^{2}$ & $0.1017084$ \\
		$\hat{\gamma}_{2}^{2}$ & $0.1841539$ \\
		$\hat{\gamma}_{3}^{2}$ & $0.3254673$ \\
		$\hat{\gamma}_{4}^{2}$ & $0.9113278$
	\end{tabular}
\end{table}

We would like to underline that the simulations of $\lambda$ are not robust. In particular, there is a very low acceptance rate in the exploited Metropolis-Hastings. We explain why this happens by an example, all the notation used in what follows, being the same as in subsection \ref{sbsec:metropolis_hastings}.

\begin{figure}[!h]
	\includegraphics[width = \columnwidth, height=.25 \paperheight ]{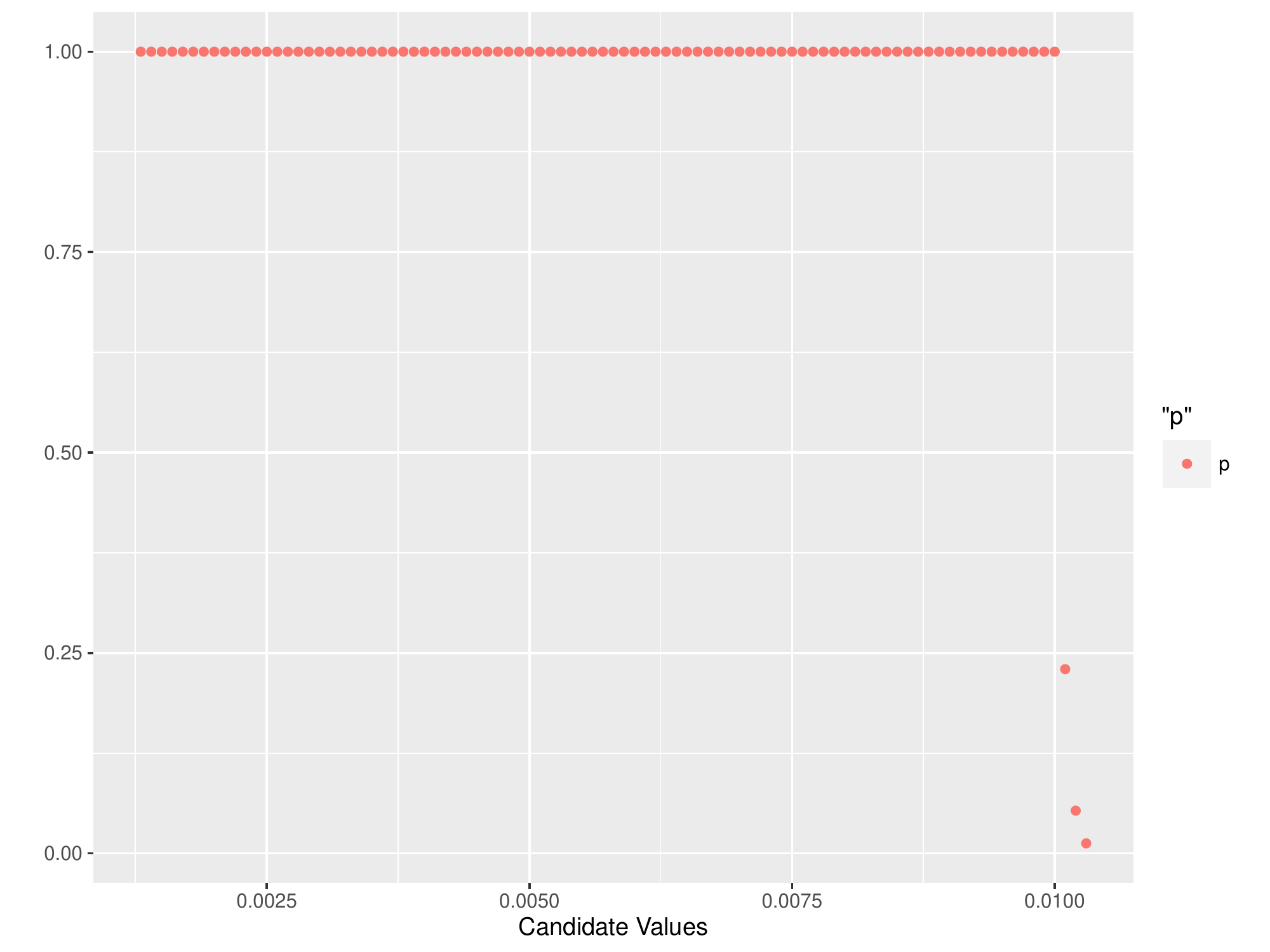}
	\caption{ \small{Acceptance probabilities of candidates around the value 0.01.} }
	\label{fig:acceptance_probabilities}
\end{figure}

We are in the special case of the Metropolis-Hastings algorithm where $\lambda = \boldsymbol \theta$ is the only parameter. If we take $\lambda^{j-1}=0.01$, where $\lambda^{j-1}$ stands for the $(j-1)$-st sample of $\lambda$, hence it is not its $(j-1)$-st power, and plot the graph of the \emph{nonzero acceptance probabilities} around that value, see Fig. \ref{fig:acceptance_probabilities}, we clearly see where the  simulations fail to be robust. 
In particular, if $\lambda_{new} \in \{0.013,0.0014,...,0.0099\}$, then its acceptance probability is 1, namely we automatically take $\lambda^{j}=\lambda_{new}$. There are only three values that $\lambda_{new}$ can take and that are larger than $0.01$, i.e.  $0.0101,\ 0.0102,\ \text{and}\ 0.0103$, with acceptance probabilities $0.23,\ 0.053,\ \text{and}\ 0.012$, respectively. The latter implies that the samples of $\lambda$ will converge towards zero, which poses a numerical problem since then the values of $\gamma_{S_{t}}$ will blow up, since  $V(y_{t}|\lambda) = \lambda \gamma_{S_{t}}^{2}$, in order to compensate. At some point the values of $\gamma_{S_{t}}$ become too large to be numerically handled, subsequently the simulation crashes. We can get around small values for $\lambda$, e.g. bounding it from below by some small constant, but then, once $\lambda$ reaches such a  value, the algorithm rarely  accept a larger value as a sample, hence leading to the problem of too few samples of $\lambda$ being accepted. Despite the aforementioned shortcoming, the proposed model still works quite effective, as we will see further down. 

\begin{figure}[!h]
	\includegraphics[width = \columnwidth, height=.25 \paperheight]{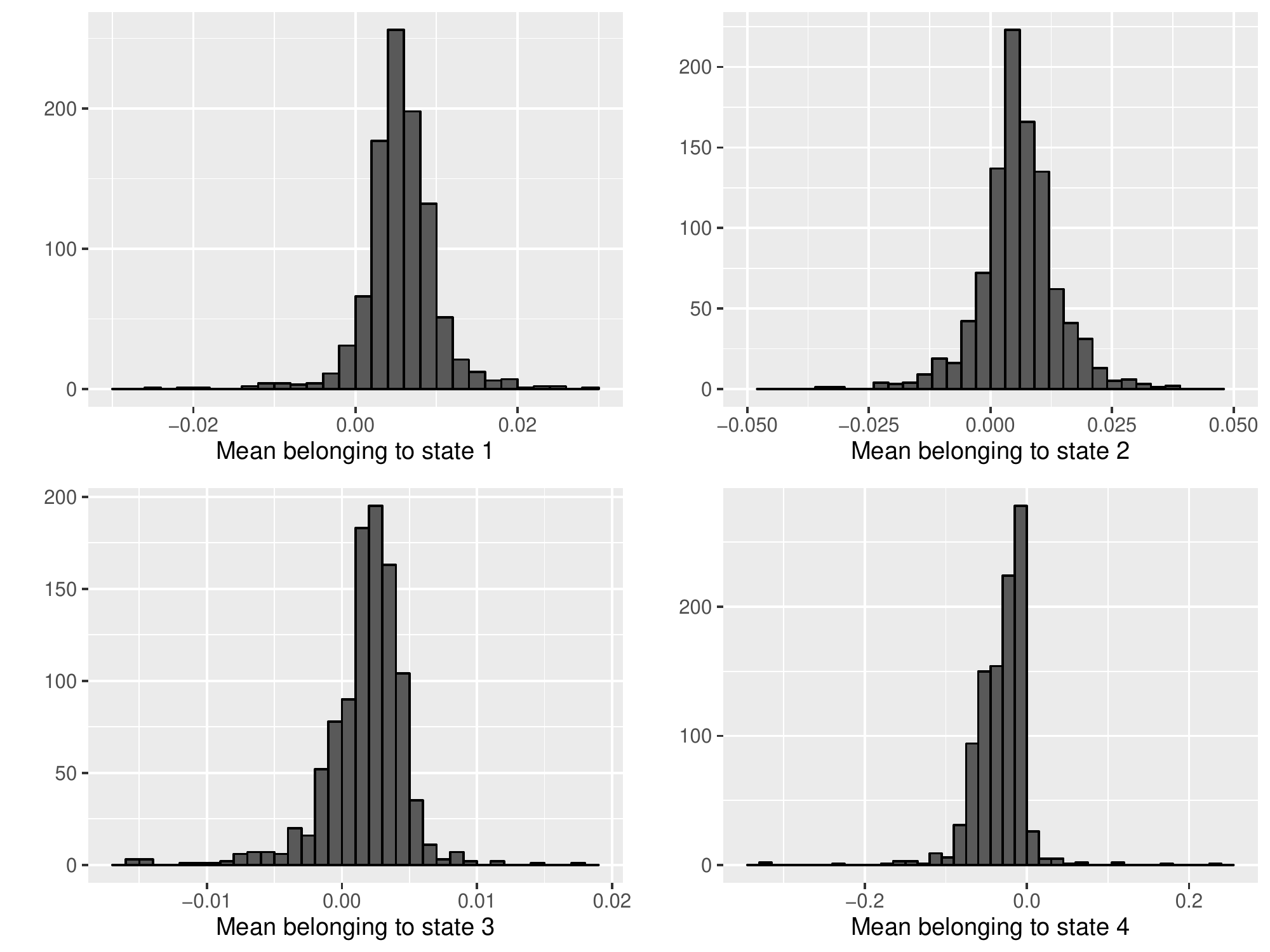}
	\caption{ \small{The histograms of the means $\mu_{S_{t}}$.} }
	\label{fig:mean_histograms}
\end{figure}

In Fig. \ref{fig:mean_histograms}, we present the histograms of the different means, while Table \ref{tbl:means} reports their point estimates.
\begin{table}[!h]
	\caption{Mean Point Estimates}
	\centering
	\begin{tabular}{c|c}
		\label{tbl:means}
		Estimator & Value \\
		\hline
		$\hat\mu_{1}$ & $0.007381561$ \\
		$\hat\mu_{2}$ & $0.008574574$ \\
		$\hat\mu_{3}$ & $0.002191889$ \\
		$\hat\mu_{4}$ & $-0.02942247$
	\end{tabular}
\end{table}

One thing that stands out in the mean point estimates, is the sign of the mean of the fourth state, which is negative. The latter should not come as a surprise since it refers to highest volatility value in the time series, namely the one related to the mortgage crisis of 2008. We recall that, during a severe financial crisis, most price movements are downward, resulting in a negative drift.

In what follows, we present the transition probability matrix, see eq. \eqref{eq:probability_matrix_stable}, and the transition matrix, see eq. \eqref{eq:transition_matrix_jump}, namely

\begin{equation}
\label{eq:probability_matrix_stable}
\hat P = \begin{bmatrix}
0.3492 & 0.2719 & 0.2102 & 0.1686\\
0.3655 & 0.2728 & 0.1993 & 0.1622\\
0.0157 & 0.0144 & 0.9518 & 0.0179\\
0.0264 & 0.0222 & 0.0220 & 0.9292
\end{bmatrix}
\end{equation} 

\begin{equation}
\label{eq:transition_matrix_stable}
\hat M = \begin{bmatrix}
29.10 & 15.65 & 7.75 & 3.62\\
16.16 & 13.76 & 5.82 & 2.95\\
6.61 & 5.54 & 270.39 & 4.17\\
4.02 & 3.49 & 3.22 & 39.68
\end{bmatrix}
\end{equation} 

\begin{figure}[!h]
	\includegraphics[width = \columnwidth, height = .25\paperheight]{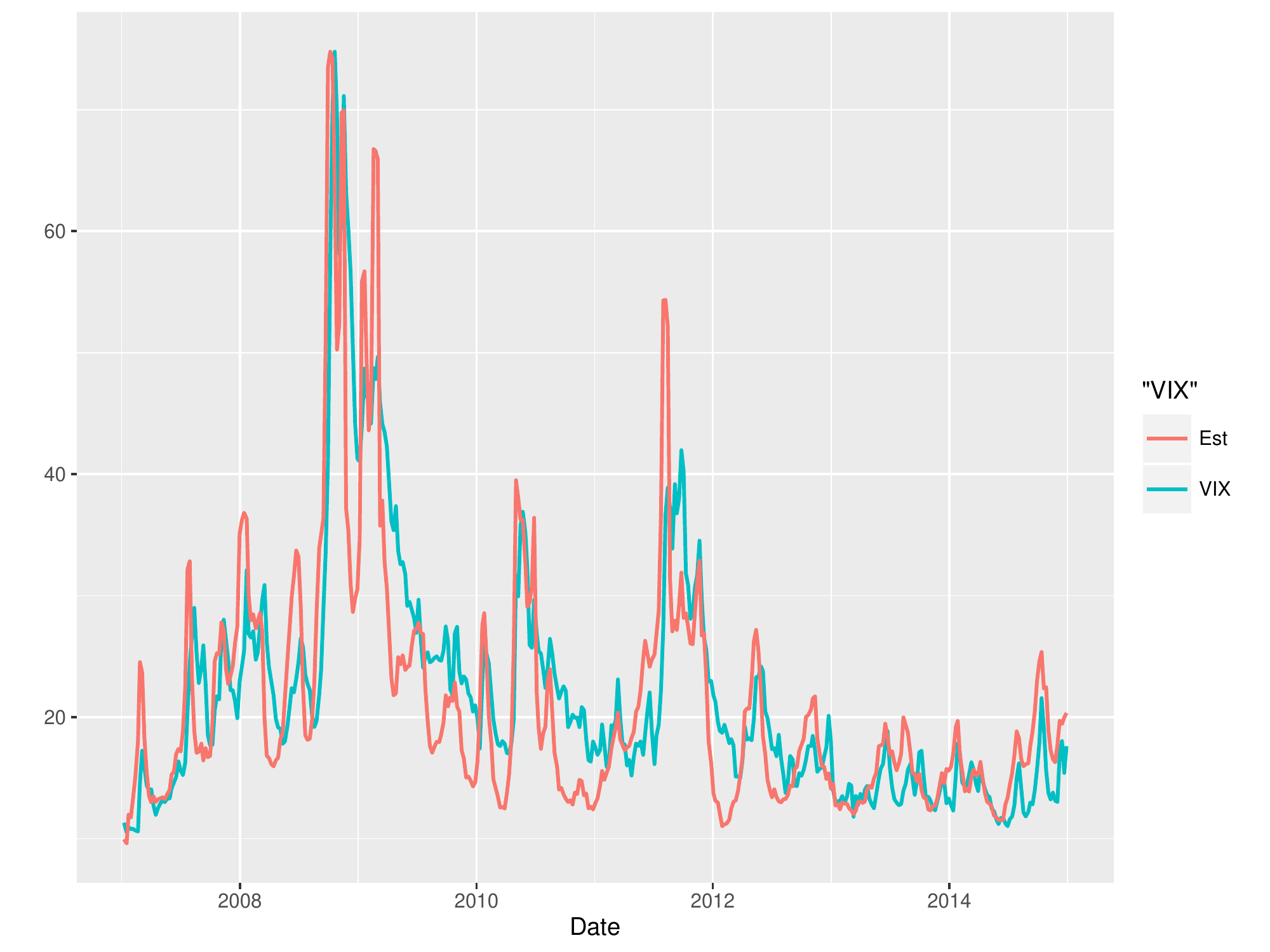}
	\caption{ \small{Visual comparison between the VIX and the volatility indicator $I_{t}^{\alpha}$. We have applied a linear scaling function to $I_{t}^{\alpha}$.} }
	\label{fig:stable_comparison}
\end{figure}

As in the previous case, we note the expected values of the state durations obtained using eq. \eqref{eq:expdur}, namely
\begin{align}
\label{duration2}
\hat{d_1} &= 1.5365, \quad \hat{d_2}=1.3751\;,\\
\nonumber \hat{d_3}&=20.746, \quad \hat{d_4}=14.124\;.
\end{align}
We can note how the difference between the results in \eqref{duration2} and those in \eqref{duration1}, is significant. In order to better  explain the latter datum, let us define the volatility indicator within the present framework, and make a comparison with the VIX index. In particular we define a second volatility indicator, denoted by $I_{t}^{\alpha}$, which, analogously to the previous case, will stand for the expected standard deviation of the data at time $t$, i.e.
\begin{equation}
I_{t}^{\alpha} = \sqrt{ \hat{\lambda} \sum\limits_{j=1}^{4} \prob (S_{t}=j|\psi_{t}) \hat{\gamma}_{j}^{2} }  \;.
\end{equation} 
In Fig. \ref{fig:stable_comparison}, we can see a visual comparison between the two values.
Using eq. \eqref{eq:squares1}, we obtain $\mathbf{S}^{\alpha}=39.69$, which is a significant increase over $\mathbf{S}^{J}$. This leads us to conclude that the estimate obtained from the jump diffusion model is closer to the VIX than the one obtained from the $\alpha$-stable distribution model.

We now briefly explain the difference between the results in \eqref{duration1} and \eqref{duration2}. The expected duration of state 1 falls while at the same time there is a drastic increase in the expected durations of states 3 and 4. Looking at the way the estimators behave in Fig. \ref{fig:jump_comparison} and in Fig. \ref{fig:stable_comparison},  it is to note that the estimator obtained from the jump diffusion model is much more {\it jagged}, because of the regular transition from one state to another; while the one obtained from the stable distribution model is much smoother, seeing as the time series tends to stay in the high volatility states much longer. Furthermore, when the stable distribution model has to place observations, that should be in the low volatility states, in the high volatility ones, so that to solve the  variance underestimation problem mentioned in \cite{DiPMF16}, the jump diffusion model can simply add a few jumps to make up for the missing variance. This is why, despite its attempts to increase the variance by staying in the higher states, we see the indicator of the stable model {\it drooping} and underestimating the low volatility, while, in this situation, the jump model stays much closer to the VIX.

\section{Conclusion and Future Developments}
In the present paper we have presented  two novel techniques to implement a   Markov Switching Model  (MSM) type approach to  non-stationary data,  namely  a jump diffusion-MSM and an $\alpha-$stable-MSM. 
In Sec. \ref{jump_diffusion_case}, we have shown that the first one is  very effective in mimic the VIX index, moreover its implementation can be smoothly done without sacrificing its theoretical peculiarities, see Sec. \ref{JumpDiffusionModel}.
A slightly different situation concerns the implementation of the second approach, see Sec. \ref{alpha_stable_case}, since eeven if the  $\alpha$-stable-MSM approach turns to be quite effective, we have to consider sampling problems of one of its parameters, implying that computational results do not  behave exactly the way the are meant to. 

We would like to underline that the achieved tractability of the jump diffusion model is a crucial point, and it  witnesses how such technique can be fruitfully used   to model any kind of time series presenting pronounced tails, not just financial ones.

As far as the  issues of over-smoothing and excessive state duration, which have been stated in  \cite{DiPMF16} as the main deficiencies of the MSM approach to financial data,  we have shown that, using the models here presented, the state durations can been significantly reduced, see subsection \ref{jump_diffusion_case}, and the problem of over-smoothing can be solved, see subesctions \ref{jump_diffusion_case} and \ref{alpha_stable_case}.
Concerning future developments, we aim at improving our jump diffusion-MSM model by considering , instead of a simple first-order Markov transition law,  a $k$-th order Markov transition law. Other possibilities consist in dealing with a transition law that is state duration dependent,  or allowing the law to depend on other observable quantities used as indicators  of the economy behaviour, e.g., real personal income, industrial production index, rate of private credit growth, etc. 
\linebreak \\
\textbf{Acknowledgements:} We would like to sincerely thank Matteo Frigo for his insightful comments and his fundamental suggestions, which have helped us a lot in preparing the present work, especially with respect to the duration analysis.





\end{document}